\documentclass[twoside]{sn-jnl}
\usepackage{graphicx}%
\usepackage{multirow}%
\usepackage{amsmath,amssymb,amsfonts}%
\usepackage{amsthm}%
\usepackage[title]{appendix}%
\usepackage{xcolor}%
\usepackage{textcomp}%
\usepackage{manyfoot}%
\usepackage{booktabs}%
\usepackage{listings}%
\usepackage[symbol]{footmisc}
\usepackage{lmodern}
\usepackage{siunitx}
\usepackage{nameref}
\usepackage{orcidlink}
\usepackage{tablefootnote}

\usepackage[sort&compress,square,numbers]{natbib}

\usepackage{etoolbox}   

\usepackage[normalem]{ulem} 

\begin{document}

\title[Emergence of a neutrino flux above 5 PeV and implications for ultrahigh-energy cosmic rays]{Emergence of a neutrino flux above 5 PeV and implications for ultrahigh-energy cosmic rays}

\author*[1]{\fnm{Marco S.} \sur{Muzio}~\orcidlink{0000-0003-4615-5529}
}\email{muzio@wisc.edu}

\author[1]{\fnm{Tianlu} \sur{Yuan}~\orcidlink{0000-0002-7041-5872}
}

\author[1]{\fnm{Lu} \sur{Lu}~\orcidlink{0000-0003-3175-7770}
}

\affil[1]{\orgdiv{Department of Physics and Wisconsin IceCube Particle Astrophysics Center}, \orgname{University of Wisconsin}, \orgaddress{\city{Madison}, \state{WI}, \postcode{53706}, \country{USA}}}

\abstract{The rare detections of astrophysical neutrinos with energies above \SI{5}{\peta \eV} by two neutrino telescopes underscore the existence of a flux at these energies. In addition to over a decade of data taken by the IceCube Neutrino Observatory, the KM3NeT neutrino telescope has recently highlighted their discovery of a possible $\mathcal{O}(\SI{100}{\peta \eV})$ neutrino candidate. A connection between the highest-energy astrophysical neutrinos and the highest-energy cosmic rays is expected, and well-established theoretically. Here, for the first time, we simultaneously fit the neutrino data from IceCube and KM3NeT, as well as the ultrahigh-energy cosmic ray spectrum and composition data from the Pierre Auger Observatory (Auger), to test a common-origin hypothesis. We show that a phenomenological model is able to describe the combined data across these three observatories, and, depending on the true energy of the event detected by KM3NeT, suggests an additional cosmic ray source population not yet robustly detected by Auger. Although a measurement of the neutrino flux in this energy regime is at the sensitivity limit of cubic-kilometer-scale neutrino telescopes, next-generation observatories, such as IceCube-Gen2, will have the sensitivity to make a significant detection of this flux.}


\maketitle

\section*{Main}\label{sec:main}

The origin of ultrahigh-energy cosmic rays (UHECRs) remains one of the enduring mysteries in astrophysics, even after more than a century of research. Recent evidence now suggests that the atomic mass of cosmic rays becomes increasingly heavy at the highest energies, indicating the presence of a mixed composition~\cite{PierreAuger:2023kjt,PierreAuger:2023yym}. This significantly complicates efforts to trace their origins due to deflections of charged particle trajectories in Galactic and extragalactic magnetic fields. For this reason, no sources of UHECRs have yet been directly identified.

Traditional gamma-ray astronomy also faces limitations in the study of UHECRs. High-energy photons can interact with the cosmic microwave background (CMB) and extragalactic background light (EBL) through pair production, significantly reducing their mean free path. These interactions prevent them from being traced directly back to their sources, and challenge efforts to identify UHECR origins through conventional observational methods.

In contrast, neutrinos, produced in the decays of subatomic particles, only interact weakly and can travel from their production sites to Earth almost entirely unimpeded. As the neutrino and UHECR horizons differ, detected UHECRs do not necessarily share the same sources as observed neutrinos~\cite{IceCube:2015afa}. Due to a rapidly decreasing flux as a function of energy, the detection of high-energy astrophysical neutrinos requires gigaton-scale detectors~\cite{Halzen:1989pva}. Neutrino telescopes, therefore, use natural media as targets, and rely on a sparse array of photomultiplier tubes to detect Cherenkov radiation produced by charged particles following neutrino interactions with the medium. Both natural bodies of water and glacial ice have been tested and employed for this function. In particular, the IceCube Neutrino Observatory has been taking data for over a decade at the South Pole~\cite{Aartsen:2016nxy}, while the KM3NeT Astroparticle Research with Cosmics in the Abyss (ARCA) neutrino telescope is under construction in the Mediterranean Sea~\cite{KM3NeT:2024paj} and the Baikal-GVD neutrino telescope is under construction in Lake Baikal~\cite{Zaborov:2024jny}. Both of the latter observatories have been taking data in partial configurations. At the highest energies, neutrino detection is also possible through the detection of extensive air showers (see e.g.~\cite{PierreAuger:2023pjg}) and radio pulses~\cite{Schroder:2016hrv}. The Pierre Auger Observatory has been searching for neutrinos with this technique for nearly two decades~\cite{PierreAuger:2023pjg}.

\par
Neutrinos beyond \SI{10}{\peta \eV} require primary particles with sufficient energy to produce them, and motivate a connection to extragalactic CRs (see, e.g.,~\cite{Fang:2017zjf,Globus:2017ehu, Kachelriess:2017tvs,AlvesBatista:2018zui,Supanitsky:2018jje,Heinze:2019jou} among others). Modern UHECR observatories, the Pierre Auger Observatory (Auger) and the Telescope Array (TA), have robustly measured both the UHECR spectrum up to $ \SI[parse-numbers = false]{\sim 10^{20.3}}{\eV}$ and composition up to $\SI[parse-numbers = false]{{\sim}10^{19.8}}{\eV}$~\cite{Anchordoqui:2018qom}. These measurements have revealed several spectral features: a spectral hardening at $\SI[parse-numbers = false]{{\sim}10^{18.7}}{\eV}$ (the \textit{ankle}), a spectral break at $\SI[parse-numbers = false]{{\sim}10^{19}}{\eV}$ (the \textit{instep}~\cite{PierreAuger:2020kuy}), and the onset of an apparent cutoff at $\SI[parse-numbers = false]{{\sim}10^{19.7}}{\eV}$~\cite{PierreAuger:2023wti}. Perhaps most importantly, while the exact composition is unclear due to uncertainties in the hadronic models used to interpret air shower data, the observed composition is inconsistent with a pure-proton hypothesis and, rather, supports a mixed composition~\cite{PierreAuger:2023bfx,PierreAuger:2023yym}.

Here, we use a sample of all four neutrinos above \SI{5}{\peta \eV} detected thus far towards interpreting the global multimessenger dataset, and present a measurement of the astrophysical neutrino flux beyond $5$~PeV. We also highlight the possibility of a non-negligible proton component beyond $30$~EeV. Specifically, we perform a joint fit of the neutrino data from IceCube, Auger, and KM3NeT/ARCA\footnote{Since Baikal-GVD and ANTARES~\cite{ANTARES:2025exu} have not detected any events above \SI{5}{\peta \eV}, and as they are expected to only marginally contribute to the global exposure, at the few percent level or less, they have a negligible impact on this analysis.} along with the latest UHECR spectrum and $X_\mathrm{max}$ distribution data from Auger to a physically motivated model of UHECR sources. We present the results under different scenarios of the true neutrino energy of KM3-230213A, and discuss implications for UHECR modeling and next-generation neutrino telescopes. Details on the methodology of the joint fits are given in the \nameref{sec:methods}.

\section*{The highest-energy events from IceCube and KM3NeT}
\label{sec:nudat}

The astrophysical neutrino flux has been measured by IceCube starting from around \SI{10}{\tera \eV} and extending up to around \SI{1}{\peta \eV}~\cite{IceCube:2020wum,IceCube:2020acn,Abbasi:2021qfz}. Above \SI{\sim5}{\peta \eV}, however, IceCube has discovered only three events in independent samples built from different selection routines. Recently, the KM3NeT/ARCA detection of a $\mathcal{O}(\SI{100}{\peta \eV})$ neutrino~\cite{KM3NeT2024,KM3NeT:2025npi} brings the total number of events at these energies to four. The handful of detections thus far can already provide some constraints on the neutrino flux while also informing UHECR models. However, larger next-generation detectors~\cite{Aartsen:2020fgd} are likely required to make a precise flux measurement above \SI{5}{\peta \eV}. Up to now, there has not been a combined analysis of the $\SI{5}{\peta \eV}$ and higher neutrino data.

\refstepcounter{footnote}\label{table_footnote} 
\begin{table}[ht]
\caption{Summary of neutrino dataset livetimes and properties of the four highest-energy neutrino events assumed in the combined fit to Auger. Note that the zenith angle, $\theta_\nu$, is given in the local-detector coordinate system with \ang{90} corresponding to the horizon, and that the inferred neutrino energy, $E_\nu$, can have a dependence on the spectrum and flavor. We consider events identified in the northern track (NT), high-energy starting events (HESE), and the PeV partially-contained events (PEPE) IceCube samples. For these events, we chose the most probable values based on available data. For KM3-230213A, the energy of which is less well constrained, we consider the published best-fit value of \SI{220}{\peta \eV}, and two values, \SI{100}{\peta \eV} and \SI{1000}{\peta \eV}, that span a wide range of possible energies based on Refs.~\cite{KM3NeT2024,KM3NeT:2025npi}. The exposures from two Auger datasets, the downgoing (DG) and the Earth skimming (ES) samples, which have not yet identified any neutrinos, are also included.}\label{tab:nuDatasets}
\begin{tabular}{@{}lrrrr@{}}
\toprule
Dataset & Livetime [yrs]  & $E_\nu$\footref{table_footnote} [PeV] & Flavor\footref{table_footnote} & $\theta_\nu$\footref{table_footnote} [deg] \\
\midrule
IceCube NT~\cite{Abbasi:2021qfz,Naab:2023xcz}    & $11.8$ & $8.7$  & $\mu$ & $101$  \\
IceCube HESE~\cite{IceCube:2013low,IceCube:2020wum,IceCube:2023sov}    & $11.8$ & $13.0$   & $\mu$  & $68$  \\
IceCube PEPE~\cite{IceCube:2021rpz}    & $4.6$ & $6.3$  & $e$  & $74$  \\
KM3NeT/ARCA~\cite{KM3NeT:2025npi} & $0.8$ & $\in \{100, 220, 1000\}$  & $\mu$ &  $89$ \\
Auger DG~\cite{PierreAuger:2019azx} & $18.0$ & – & – & – \\
Auger ES~\cite{PierreAuger:2019azx} & $18.0$ & – & – & – \\
\botrule
\end{tabular}
  \begin{tablenotes}
    \item[\footref{table_footnote}] Assumed in this work
  \end{tablenotes}
\end{table}

Table~\ref{tab:nuDatasets} lists KM3-230213A along with three IceCube events with inferred neutrino energies above \SI{5}{\peta \eV}. The ``Dataset'' column indicates the corresponding sample in which the events were discovered. For IceCube, these correspond to three independent selections: the northern track (NT) sample~\cite{Abbasi:2021qfz,Naab:2023xcz}, the high-energy starting events (HESE) sample~\cite{IceCube:2013low,IceCube:2020wum,IceCube:2023sov}, and the PeV partially-contained events (PEPE) sample~\cite{IceCube:2021rpz}. Using different techniques, each approach succeeds in isolating a neutrino signal from the substantial down-going atmospheric muon background. The NT sample focuses on selecting muon-induced tracks from the northern sky, primarily from charged current muon neutrino interactions outside the IceCube instrumented volume. The HESE sample targets all-flavor neutrino interaction within a fiducial region of the detector, whereby only events were selected that produced more than $6000$~photoelectrons in the PMTs. The PEPE sample is specifically tuned for sensitivity to Glashow resonance~\cite{PhysRev.118.316} detection and expands the effective volume by selecting for electron- and tau-neutrino interactions, as well as muon-neutrino neutral current interactions, occurring near or outside the instrumented region. There is negligible overlap among the three IceCube samples due to their differing signal topology requirements. The PEPE selection was developed to be complementary to HESE, resulting in approximately \SI{1}{\%} overlap above \SI{5}{\peta \eV}. Both selections reject throughgoing tracks such as those included in the NT sample~\cite{Lu:2017nti}. We assume the latest livetimes of the three samples. For each event, the most-probable neutrino energy, flavor and zenith angle\footnote{The zenith angle corresponds to the polar angle of the neutrino arrival direction as defined in local detector coordinates; $\theta_\nu=0$ corresponds to a neutrino entering from directly above the detector.} is taken. While this ignores reconstruction resolutions, such an approximation is reasonable given the large energy and zenith bins used in the binned likelihood construction (c.f.~\nameref{sec:methods}).

KM3-230213A\footnote{That this event was not observed by IceCube has generated significant discussion in the literature. Several other studies have estimated the level of tension between this observation by KM3NeT and non-observation by IceCube~\cite{Li:2025tqf,KM3NeT:2025ccp}. Given its large energy uncertainty, however, the focus of this paper is on the astrophysical implications of this observation, assuming a few plausible neutrino energies for KM3-230213A.} is the highest-energy neutrino candidate event first announced in Ref.~\cite{coelho_2024_12706075} and published in Ref.~\cite{KM3NeT:2025npi}. ARCA has operated in a partially constructed configuration of 6, 8, 19 and 21 detection units (DU), with their highest-energy event detected under the 21 DU configuration. For this analysis, we approximate the relevant livetime to be roughly that of the combined ARCA19-21 data-taking periods, or about 0.8 years~\cite{KM3NeT:2025npi}. Assuming an $E^{-2}$ spectrum, the neutrino energy of the event is estimated to be \SI[parse-numbers = false]{220^{+570}_{-110}}{\peta \eV}~\cite{KM3NeT:2025npi}. As the uncertainties are large and the energy estimate depends on the spectrum and flavor, we test three different neutrino energy hypotheses spanning this range: \SI{100}{\peta \eV}, \SI{220}{\peta \eV}, and \SI{1}{\exa \eV}.

Finally, we also include the exposure from two Auger datasets: the downgoing (DG) sample and the Earth skimming (ES) sample~\cite{PierreAuger:2019azx}. Neither of these datasets have identified neutrino candidates as of the latest analysis~\cite{PierreAuger:2023pjg}.

To perform a maximum likelihood analysis, we use the published IceCube, Auger, and KM3NeT/ARCA effective areas for each sample. Due to instrumentation effects and neutrino Earth absorption, the effective areas are dependent on $(E_\nu, \theta_\nu, \mathrm{flavor})$ as well as the event selection itself. A product of flux, livetime and effective area gives the number of expected astrophysical neutrinos of a given flavor in each sample at a corresponding energy and zenith angle. Since the samples are independent, a simple sum over their individual log-likelihoods assuming Poisson statistics can be performed. Finally, we assume that backgrounds are negligible above \SI{5}{\peta \eV}, with atmospheric muons rejected by some combination of overburden or veto~\cite{IceCube:2020wum,Abbasi:2021qfz}, and atmospheric neutrino rates suppressed by accompanying muons~\cite{Arguelles:2018awr,IceCube:2020wum} and model expectations~\cite{Workman:2022ynf,Abbasi:2021qfz}.

\section*{Ultrahigh-energy cosmic rays and the Pierre Auger Observatory}

\par
The flux of UHECRs is extremely low, on the order of ${\sim}1$ per $\text{km}^2$ per century at the highest energies ($\sim 10^{20}$ eV). Detecting them requires observatories with effective collection areas of several thousands of square kilometers. Direct cosmic-ray detectors in space, such as AMS-02, are limited by their small collection areas (a few square meters) and cannot effectively detect UHECRs,  despite their ability to measure the charge and mass of lower-energy cosmic rays. Instead, UHECRs can only be detected indirectly through the extensive air showers they initiate in Earth’s atmosphere~\cite{Workman:2022ynf}. These particle showers are observed on the basis of secondary particles arriving at ground-based particle detectors, as well as through fluorescent emission in the atmosphere as the shower develops. Modern UHECR observatories leverage both of these observables by combining data from fluorescence telescopes and a ground based array of either scintillator panels or water Cherenkov tanks. Additionally, these detectors can be augmented by the addition of a radio antenna array. The energy of the primary particle can be robustly estimated from these observables using well-known techniques~\cite{PierreAuger:2021hun}. However, the primary particle's chemical composition must be inferred from the energy and depth of shower maximum, $X_\mathrm{max}$, and requires modeling hadronic interactions at ultrahigh energies. While this cannot be done on an event-by-event basis, analysis of the distribution of $X_\mathrm{max}$ values at a given energy allows the relative abundances of different nuclei to be inferred. Typically, the first two moments of these distributions are used as input to fit the composition of the UHECR observations. In this work, we instead fit the full $X_\mathrm{max}$ distributions in order to take advantage of the additional information encoded in higher-order moments.

\begin{figure}[htbp]
    \centering
    \includegraphics[width=0.9\textwidth]{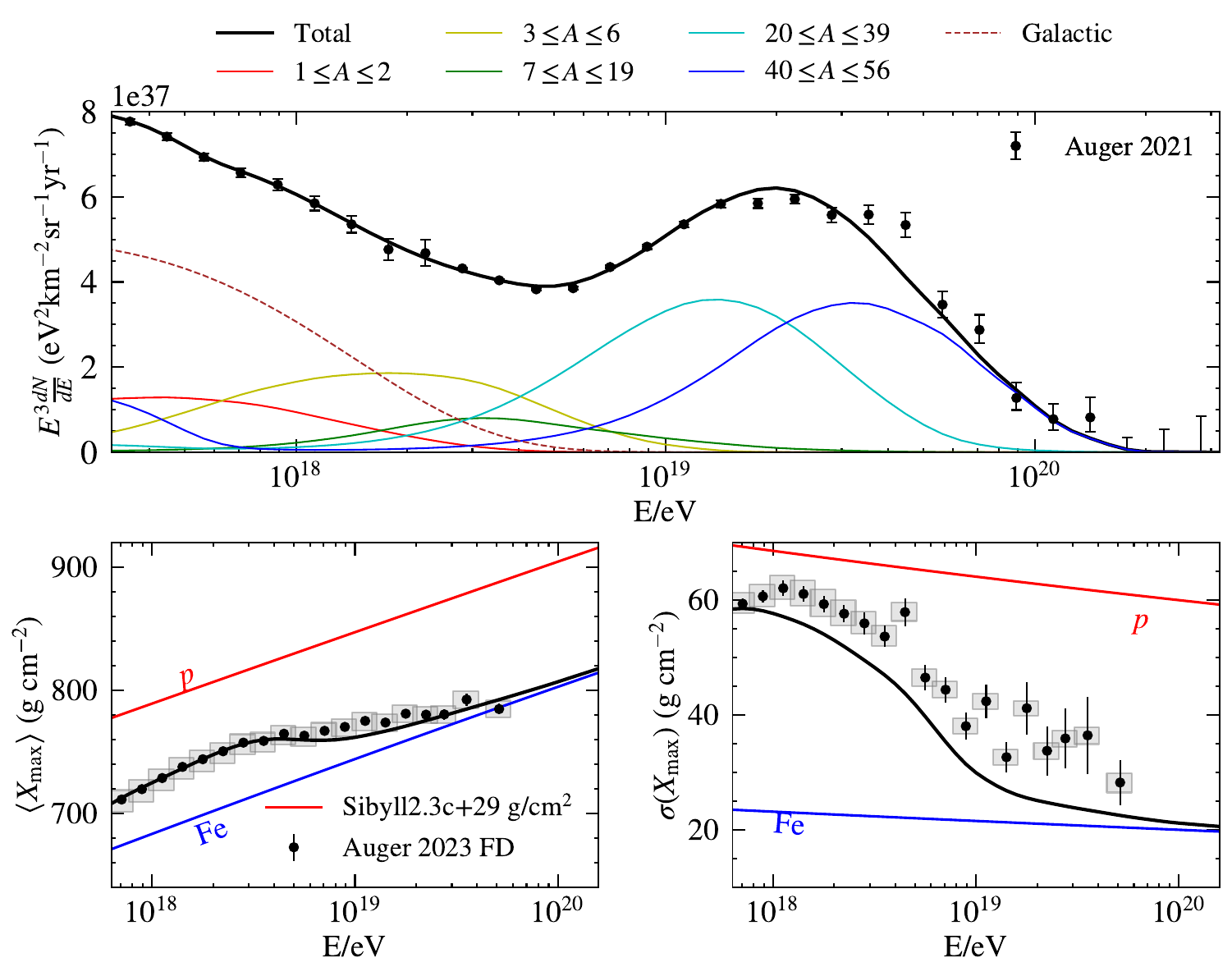}
    \caption{The measured and predicted CR spectrum (top panel) and $X_\mathrm{max}$ moments (bottom panels) for the best-fit model assuming KM3-230213A has an energy of \SI{100}{\peta \eV}, using the joint likelihood. The total spectrum (black line) is plotted along with the contribution from each mass group and Galactic CRs (colored lines), as compared to observations from Auger~\cite{PierreAuger:2021hun}. Unfolded $\langle X_\mathrm{max} \rangle$ and $\sigma(X_\mathrm{max})$ measurements from Auger~\cite{PierreAuger:2023kjt} are plotted along with model predictions interpreted using the \textsc{Sibyll2.3c} hadronic interaction model (shifted by $+29$~g/cm$^2$) and its expectation for pure proton (red line) and Fe (blue line) compositions.}\label{fig:fid_CRspec}
\end{figure}

In this work we include two Auger datasets in our joint likelihood. The first is the latest measurement of the CR spectrum above \SI[parse-numbers = false]{10^{17}}{\eV}~\cite{PierreAuger:2021hun}. This unfolded spectrum combines 15 years of air shower observations from Auger's two water Cherenkov detector arrays: the \SI{750}{\m} array and the \SI{1500}{\m} array, encompassing roughly \SI{24}{\kilo \m^2} and \SI{3000}{\kilo \m^2}, respectively. 

\par
In addition, we fit the CR composition via the latest published $X_\mathrm{max}$ distributions, which have not been corrected for detector effects, including the detector acceptance and resolution, which can broaden and bias these distributions~\cite{PierreAuger:2023kjt}. This dataset combines roughly 17~years of data from Auger's 24~fluorescence telescopes, along with observations from three specialized fluorescence telescopes for high-elevation measurements. For visualization purposes, we present the results produced from this dataset as unfolded $\langle X_\mathrm{max} \rangle$ and $\sigma(X_\mathrm{max})$ measurements. We note that Auger has also recently published $\sigma(X_\mathrm{max})$ measurements from surface detector data using a DNN~\cite{PierreAuger:2024flk,PierreAuger:2024nzw}, which indicate a narrower $X_\mathrm{max}$ distribution above \SI[parse-numbers = false]{10^{18.5}}{\eV} than the fluorescence detector measurements presented here. This data would provide additional constraints on, in particular, the high-energy proton component we explore below. However, given that this data infers an unphysical composition under \textsc{Sibyll2.3d} (see Figs. 16 and 17 of~\cite{PierreAuger:2024nzw}), we omit it from this work.

Figure~\ref{fig:fid_CRspec} shows this latest CR spectrum and composition data from Auger, along with model predictions for the \SI{100}{\peta \eV} scenario. The spectrum prediction (top panel) is broken down into five mass groups (representing $p$-, He-, CNO-, Si-, and Fe-like nuclei), as well as, the contribution from Galactic CRs. Air shower observations (bottom panels), in the form of the first two moments of the $X_\mathrm{max}$ distribution, are shown relative to expectations for pure proton and Fe compositions, interpreted using \textsc{Sibyll2.3c} shifted by $+29$~g/cm$^2$ based on Ref.~\cite{PierreAuger:2024neu}.

\section*{Diffuse joint fit with UHECR and neutrinos}\label{sec:results}

At UHEs, the energy of neutrinos produced by CRs interacting in extragalactic propagation (cosmogenic neutrinos) or inside their source environments (astrophysical neutrinos) is strongly correlated to the energy of the primary CR. In particular, the primary CR's energy-per-nucleon sets the energy scale of the secondary neutrino, since interactions occur off of individual nucleons due to the much larger center-of-mass energy relative to the nuclear binding energy. The energy of a photohadronically produced neutrino is roughly $E_\nu \simeq E_A/A/20$, where $E_A$ is the energy of a CR with mass number $A$~\cite{Anchordoqui:2018qom}. Neutrinos produced hadronically typically have much lower energies than those produced photohadronically, due to the much larger particle multiplicity involved. 

\par
Nuclei interactions can also generate neutrinos via photopion production. However, as the photonuclear cross-section at UHEs is dominated by the giant dipole resonance, CR nuclei are more likely to photodisintegrate, a process that results in secondary nucleons, rather than neutrinos, with approximately the same energy-per-nucleon as the parent CR~\cite{Murase:2010gj,Anchordoqui:2018qom}. Given sufficient propagation times, these secondary nucleons can still photopion produce neutrinos with energies comparable to those which their parent nuclei would have created. This process is particularly important for secondary nucleons produced inside source environments, where interaction times are relatively short. 

\par
The energy scale of ambient photons sets the threshold energy for CRs to photopion produce. For a target photon with energy $\omega$, this threshold energy is given by $E_A/A = \frac{m_\pi (2m_p + m_\pi)}{4\omega} \simeq 10^{19.5} \left(\frac{\omega}{10^{-3}~\mathrm{eV}}\right)^{-1}$~eV~\cite{Anchordoqui:2018qom}. Since the CR spectrum falls off rapidly, secondary neutrinos will be primarily produced by CRs at threshold. A consequence of this is that neutrinos produced via CR interactions during extragalactic propagation will result in several typical energy scales. Interactions with the CMB result in a peak neutrino flux at roughly \SI{1}{\exa \eV}, but require primary CRs with $E_A/A \gtrsim 10^{19.3}$~eV. Interactions with the EBL, in particular the cosmic infrared background (CIB), result in a peak neutrino flux at roughly \SI{30}{\peta \eV} and only require primary CRs with $E_A/A \gtrsim 10^{18}$~eV. Finally, the peak energy of neutrinos produced within source environments will correspond to roughly the lowest energy where there are significant numbers of interactions before escape.

\par
While existing CR spectrum and composition data require only a CR population with energies $\gtrsim$~\SI{1}{\exa \eV} which could photopion produce with the EBL, they do not rule out a second CR population with energies $\gtrsim$~\SI{20}{\exa \eV} which could photopion produce off of the CMB. Indeed, the long tails of the latest $X_\mathrm{max}$ distributions from Auger~\cite{PierreAuger:2023kjt} may hint at a proton contribution at the highest energies~\cite{PierreAuger:2023xfc}. These protons could originate from an additional extragalactic source population that only contributes to the observed UHECR flux at the highest energies, and therefore is not captured by the model explored in~\cite{PierreAuger:2022atd}. Protons at these energies would be strongly attenuated by interactions with the CMB, a phenomenon referred to as the Greisen-Zatsepin-Kuzmin (GZK) effect~\cite{Greisen:1966jv,Zatsepin:1966jv}. Several recent studies have explored the compatibility of such a pure-proton component at the highest energies with data and its multimessenger implications~\cite{vanVliet:2019nse,Muzio:2019leu,Anker:2020lre,Das:2020nvx,Ehlert:2023btz,Muzio:2023skc}. Notably, these studies have found that a proton component above $\SI{10}{\exa \eV}$ could make up as much as $10\%$ of the UHECR flux above \SI{50}{\exa \eV}, while improving fits to the UHECR spectrum \& composition data, and saturating current bounds on the neutrino flux above \SI{10}{\peta \eV}.

\par
The UHECR source model used in this analysis~\cite{Unger:2015laa,Muzio:2019leu,Muzio:2021zud} properly accounts for each of these neutrino production processes, as well as Bethe-Heitler energy losses and secondary CR production from photodisintegration inside source environments and in extragalactic propagation. Therefore this model generically predicts both astrophysical and cosmogenic contributions to the neutrino flux. A thorough discussion of the details of this model can be found in the \nameref{sec:methods}.

\begin{table}[htbp]
\caption{UHECR-neutrino deviances for the UHECR and SPL models under the three KM3NeT/ARCA energy scenarios. For each combination, the number of degrees of freedom $N_\mathrm{dof}$ and the total joint deviance $D$ are reported, along with the contributions from the UHECR spectrum $D_{\mathrm{CR},J}$, UHECR composition $D_{\mathrm{CR},X}$, low-energy neutrino data $D_{\nu,\mathrm{lo}}$, and high-energy neutrino data $D_{\nu,\mathrm{hi}}$. The $p$-values for the high-energy neutrino deviance of the UHECR model relative to the SPL case is also reported (see \nameref{sec:methods} for details). Each is derived assuming a specific energy for KM3-230213A, and are illustrative of a few plausible scenarios. An asterisk ($^*$) indicates the model uses an additional pure proton component at GZK energies.}\label{tab:fitSummaries}
\begin{tabular}{@{}lllllllll@{}}
\toprule
KM3NeT Energy & Model & $N_\mathrm{dof}$ & $D$ & $D_{\mathrm{CR},J}$ & $D_{\mathrm{CR},X}$ & $D_{\nu,\mathrm{lo}}$ & $D_{\nu,\mathrm{hi}}$ & $p$-value \\
\midrule
\SI{100}{\peta \eV} & UHECR model & $2809$ & $1793.6$ & $527.9$ & $1203.8$ & $18.1$ & $43.8$ & $6.7\times10^{-1}$\\
\SI{100}{\peta \eV} & SPL model    & $2037$ & $60.3$ & N/A & N/A & $18.1$ & $42.2$ & N/A \\
\specialrule{0.1pt}{1pt}{1pt}
\SI{220}{\peta \eV} & UHECR model & $2809$ & $1797.5$ & $527.0$ & $1203.9$ & $18.1$ & $48.5$ & $9.1\times 10^{-1}$ \\
\SI{220}{\peta \eV} & UHECR model$^*$ & $2806$ & $1689.4$ & $544.2$ & $1085.3$ & $18.1$ & $41.8$ & $6.6\times 10^{-2}$ \\
\SI{220}{\peta \eV} & SPL model & $2037$ & $61.6$ & N/A & N/A & $18.1$ & $43.5$ & N/A \\
\specialrule{0.1pt}{1pt}{1pt}
\SI{1}{\exa \eV} & UHECR model$^*$   & $2806$ & $1687.5$ & $546.2$ & $1081.9$ & $18.1$ & $41.3$ & $4\times10^{-4}$ \\
\SI{1}{\exa \eV} & SPL model &  $2037$ & $65.4$ & N/A & N/A & $18.0$ & $47.3$ & N/A \\
\botrule
\end{tabular}
\end{table}

In this section, we analyze the UHECR--neutrino connection using data from Auger, IceCube, and KM3NeT/ARCA. Given the energy uncertainty of the neutrino candidate observed by ARCA21, which depends strongly on the underlying neutrino spectrum and flavor, we consider three scenarios for its true energy: \SI{100}{\peta \eV}, the reported central value of \SI{220}{\peta \eV}, and \SI{1}{\exa \eV}.

\par
For each scenario we compare two descriptions of the neutrino spectrum at Earth: a standalone single power law (hereafter the SPL model), and the spectrum predicted by the joint UHECR--neutrino model (hereafter the joint model; labelled ``UHECR model'' in Table~\ref{tab:fitSummaries}). The SPL model is
\begin{align}\label{eq:SPL}
    \Phi_\mathrm{SPL}(E_\nu) = \phi_\mathrm{SPL} \left(\frac{E_\nu}{\SI[parse-numbers = false]{10^{14}}{\eV}}\right)^{\gamma_\mathrm{SPL}}~,
\end{align}
with an assumed $1:1:1$ flavor ratio, no cutoff, and two free parameters: the all-flavor normalization $\phi_\mathrm{SPL}$ at \SI[parse-numbers = false]{10^{14}}{\eV} and the spectral index $\gamma_\mathrm{SPL}$ (best-fit values in Table~\ref{tab:modelParams}). It is fit to the neutrino data alone: the IceCube flux measurements between \SI{30}{\tera \eV} and \SI{2}{\peta \eV} and the event counts above \SI{5}{\peta \eV} from the datasets in Table~\ref{tab:nuDatasets}. The joint model's neutrino spectrum is the sum of a low-energy nuisance component and a UHECR-produced component. The low-energy nuisance component is a power law with the cutoff of Eq.~\eqref{eq:CRspectrum} and a $1:1:1$ flavor ratio, with three free parameters ($\gamma_\nu$, $E_{\mathrm{max},\nu}$, and $\phi_{\nu 0}$ in Table~\ref{tab:modelParams}). The UHECR-produced component is the flux of neutrinos generated by the fitted UHECR population itself, through photohadronic and hadronic interactions inside the source environment and photohadronic interactions during extragalactic propagation~\cite{Muzio:2021zud}. Its normalization is inherited from the fit of the propagated CR spectrum to the Auger spectrum, together with the assumed source evolution (see \nameref{sec:methods}), and its spectral shape follows from the same source-environment and propagation calculation that determines the CR spectrum and composition at Earth, so that this component is fixed almost entirely by the CR data. The flavor composition of this component at Earth is also predicted by the model. Neutrinos from $\pi^\pm$ decay and $\bar{\nu}_e$ from the $\beta$-decay of neutrons are propagated to Earth with flavor-averaged oscillation probabilities~\cite{Anchordoqui:2013dnh}, which results in small, energy-dependent deviations from $1:1:1$.

Depending on the model, we fit to the data by minimizing the joint deviance $D=D_{\mathrm{CR},J} + D_{\mathrm{CR},X} + D_{\nu,\mathrm{lo}} + D_{\nu,\mathrm{hi}}$ to the UHECR and neutrino data, or $D= D_{\nu,\mathrm{lo}} + D_{\nu,\mathrm{hi}}$ to the neutrino data alone (see Table~\ref{tab:fitSummaries}). The UHECR models presented here are those with the smallest total neutrino deviance for models whose joint deviance obeys $D\leq D_\mathrm{min} + 1$, where $D_\mathrm{min}$ is the deviance of the best-fit model to the joint dataset. The corresponding parameters for these models can be found in Table~\ref{tab:modelParams} in the \nameref{sec:methods}. We quantify the statistical preference for the joint model relative to the SPL model of the same scenario in Table~\ref{tab:fitSummaries} using an empirical $p$-value based on the difference of the two models' high-energy neutrino deviances $D_{\nu,\mathrm{hi}}$. A detailed description of each contribution to the deviance and of the empirical $p$-value can be found in the \nameref{sec:methods}. Notably, the increased significance of the UHECR models over the SPL model with the KM3NeT event's energy is, in part, driven by the worsening description of that event by the SPL model itself.

We find that in the \SI{100}{\peta\eV} scenario, the observed neutrino data is consistent with a single power-law (SPL) flux, whereas the preferred UHECR model predicts a neutrino component peaking at $\sim{\SI{30}{\peta\eV}}$. In the \SI{1}{\exa\eV} scenario, we introduce an additional pure proton population at the GZK limit to explore the possibility of the first observed GZK neutrino. Finally, we consider implications of the central \SI{220}{\peta\eV} value reported by~\cite{KM3NeT:2025npi}.

\subsection*{\SI{100}{\peta\electronvolt} case: Implications for the astrophysical neutrino flux}

\begin{figure}[htbp]
    \centering
    \includegraphics[width=0.48\linewidth] {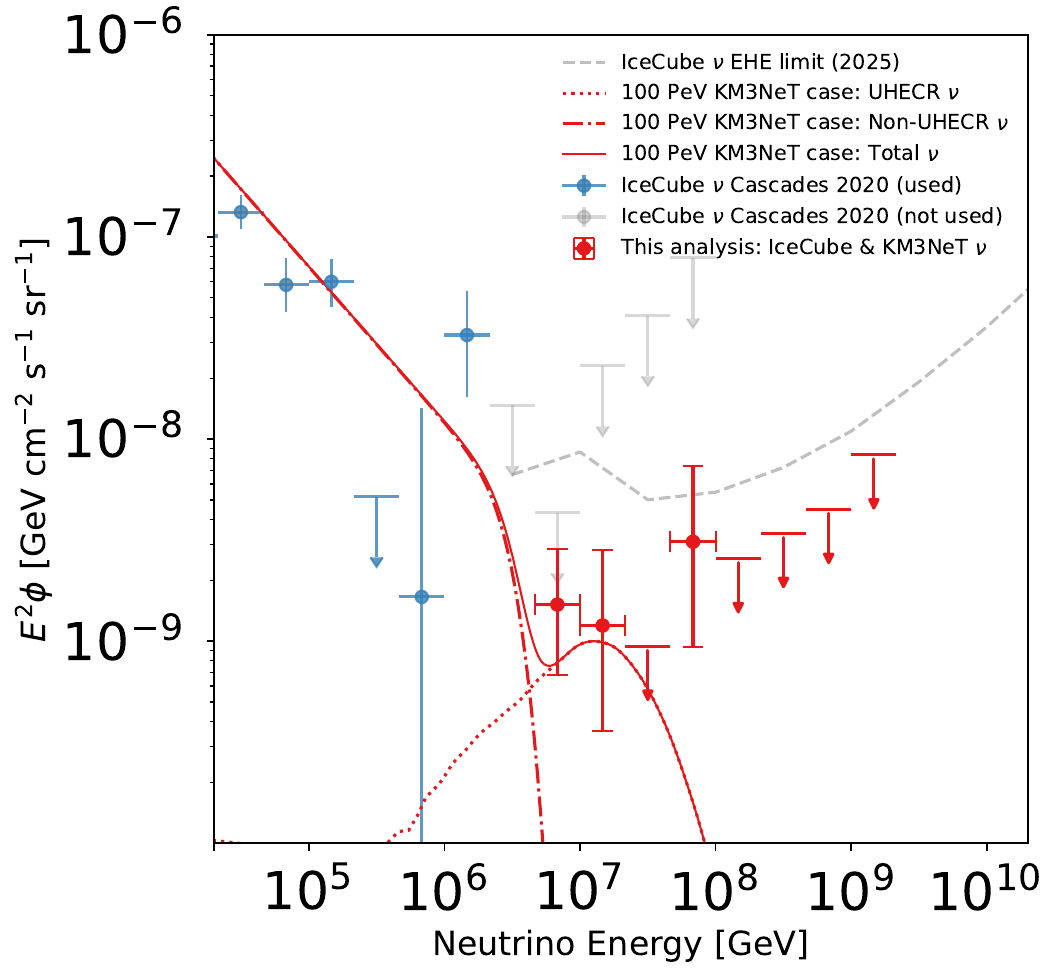} 
    \includegraphics[width=0.48\linewidth]{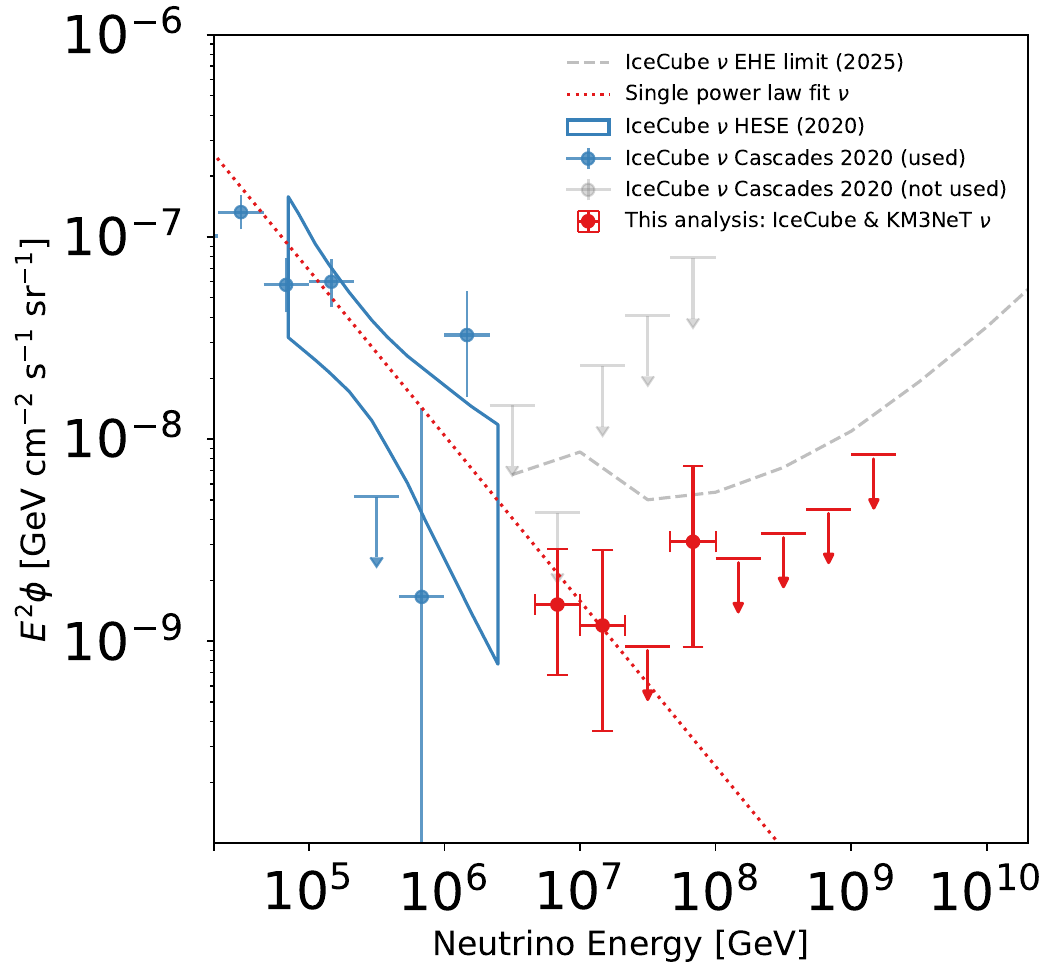} 
    \caption{Results obtained for a \SI{100}{\peta \eV} KM3NeT/ARCA neutrino. The left panel shows the predicted neutrino flux for the model that best describes the combined neutrino and cosmic ray data. The right panel shows the best-fit single-power law (SPL) flux obtained by fitting the neutrino data alone. The IceCube measured flux from Ref.~\cite{IceCube:2020acn} is shown in solid blue and gray, with the lower energy values (blue) included in the fits. Differential upper limits (90\% CL) from Ref.~\cite{IceCubeCollaborationSS:2025jbi} are shown as the dashed gray line. The red points show the segmented $E^{-2}$ flux obtained by fitting independent normalizations in each energy segment to all high-energy neutrino datasets of Table~\ref{tab:nuDatasets}, with the assumed energy of KM3-230213A for this scenario (see \nameref{sec:methods}). Error bars and arrows are $1\sigma$ profile-likelihood intervals and 84\% CL upper bounds.}
    \label{fig:mid}
\end{figure}
As a baseline scenario, we assume that the true energy of the neutrino observed by ARCA21 is \SI{100}{\peta \eV}, roughly the lower end of the $1\sigma$ energy interval reported in Ref.~\cite{KM3NeT:2025npi}. Under this assumption we consider two models: an unconstrained, single-power law (SPL) flux and a neutrino flux produced by UHECRs. The model fluxes are shown as red lines in Fig.~\ref{fig:mid}. In both cases, the data from Table~\ref{tab:nuDatasets} and the IceCube measured neutrino flux from \SI{\sim 30}{\tera \eV} to \SI{\sim 5}{\peta \eV} (blue)~\cite{IceCube:2020acn} are included in the fit. In the case of the UHECR model (left panel), Auger data is additionally included. The gray upper limits are also from Ref.~\cite{IceCube:2020acn} but are excluded from the model fits in lieu of Table~\ref{tab:nuDatasets}.

For comparison, the red points show a model-independent, segmented $E^{-2}$ flux fit to the neutrino data in Table~\ref{tab:nuDatasets} with the same likelihood as the model fits, as described in the \nameref{sec:methods}. The segmented fit is repeated for each assumed energy of KM3-230213A and is shown in Figs.~\ref{fig:mid}, \ref{fig:hi}, and \ref{fig:central}. In this context, we see the emergence of a neutrino flux at energies beyond \SI{5}{\peta \eV}.

As the right panel of Fig.~\ref{fig:mid} shows, a \SI{100}{\peta \eV} neutrino provides hints of a deviation from the SPL expectation. In contrast, as shown in the left panel, under the assumption that the neutrino flux above \SI{10}{\peta \eV} is primarily of UHECR-origin, a very different qualitative picture emerges: the SPL spectrum observed at low energies transitions to a UHECR-produced neutrino component with a recovery in the spectrum that peaks around \SI{30}{\peta \eV}. The UHECR-origin model comparably describes the high-energy neutrino data, as shown in $D_{\nu,\mathrm{hi}}$ in Table~\ref{tab:fitSummaries}, and current statistics are insufficient to distinguish between these two models above \SI{5}{\peta \eV} (as evidenced by its $p$-value of $6.7\times 10^{-1}$). However, the UHECR prediction can describe both the global neutrino data and the UHECR spectrum and composition data (see Fig.~\ref{fig:fid_CRspec}), lending it significant phenomenological evidence. We note that while here we only present the model giving the best-fit to the neutrino data within $1\sigma$ of our best-fit model, many UHECR theoretical and phenomenological models in the literature predict a significantly larger flux above \SI{5}{\peta \eV} than the SPL model, and refer the reader to Ref.~\cite{Valera:2022wmu} for a concise summary of many such models. A number of astrophysical models for specific source types, including tidal disruption events (TDEs)~\cite{Biehl:2017hnb,Plotko:2024gop} and jetted active galactic nuclei (AGN)~\cite{Rodrigues:2020pli}, also predict neutrino fluxes peaking in this energy range.

\subsection*{\SI{1}{\exa \eV} case: Implications for the cosmogenic neutrino flux}

\begin{figure}[htbp]
    \centering
    \includegraphics[width=0.48\linewidth] {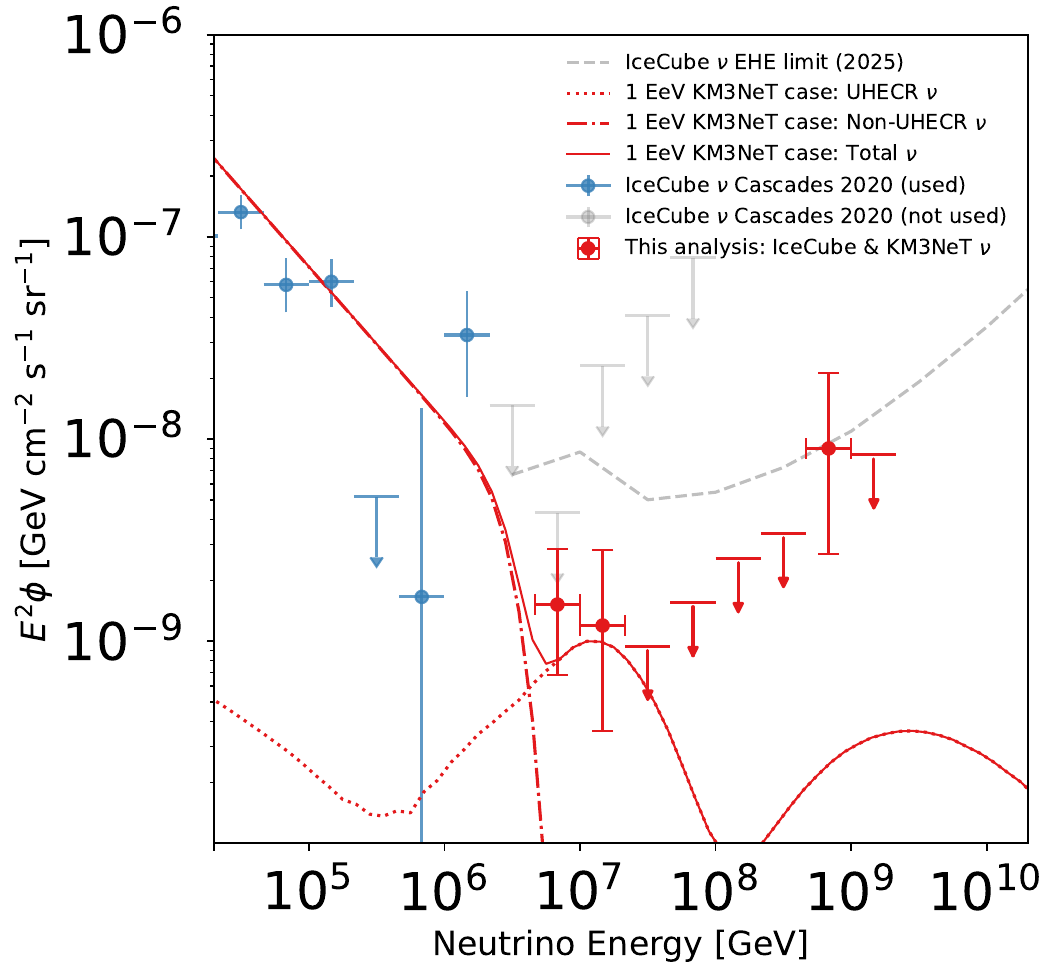} 
    \includegraphics[width=0.48\linewidth]{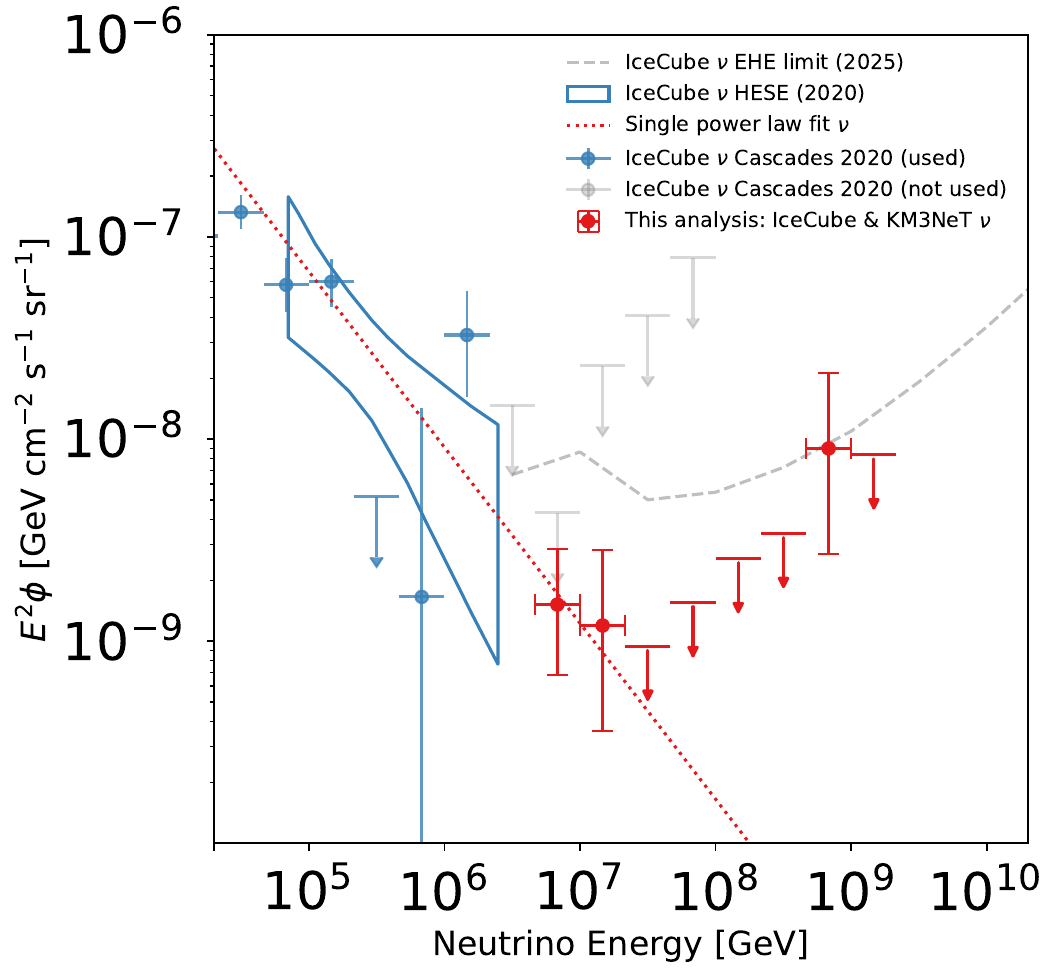} 
    \caption{Same as Fig.~\ref{fig:mid} except KM3-230213A is assumed have an energy of \SI{1}{\exa \eV}, and the UHECR model shown in the left panel includes an extra pure-proton population at the highest energies. The segmented flux (red points) is re-fit with KM3-230213A at this energy. A \SI{1}{\exa \eV} neutrino candidate makes it more difficult to reconcile the SPL model shown in the right panel with the neutrino data.}
    \label{fig:hi}
\end{figure}

It is also possible that the true neutrino energy is significantly larger than the \SI{100}{\peta \eV} assumed in the previous section -- even as high as \SI{1}{\exa \eV}. In order to produce such an energetic neutrino, its parent CR would need to have $E_A / A \approx \SI{20}{\exa \eV}$. The UHECR model presented in the previous section had a maximum rigidity (energy-per-charge) of less than \SI{10}{\exa \eV}, in agreement with the results presented in Ref.~\cite{PierreAuger:2022atd}. For nuclei this translates to a maximum $E_A / A < \SI{5}{\exa \eV}$. Thus, the baseline UHECR source population assumed earlier is unlikely to produce such an energetic neutrino observation.

A second UHECR source population, driven by a more powerful accelerator, would naturally motivate such an observation. To test this, we include an additional proton component that escapes its source environment with a maximum rigidity above \SI{10}{\exa \eV}. The compatibility of such a two-component scenario with UHECR data has been investigated by several previous studies using existing UHECR composition data~\cite{vanVliet:2019nse,Muzio:2019leu,Anker:2020lre,Ehlert:2023btz,Muzio:2023skc}. Such a model was also considered by~\cite{KM3NeT:2025vut} in the context of the KM3NeT observation. However, in that study, they did not attempt to fit the global neutrino data. Here, we show that a two-component model could explain a $E_\nu = \SI{1}{\exa \eV}$ KM3NeT/ARCA observation, and explore its implications for UHECR detectors.

Figure~\ref{fig:hi} shows the predicted neutrino flux for the best-fit two-component UHECR source model (left panel) and the best-fit SPL neutrino flux (right panel) assuming an $E_\nu = \SI{1}{\exa \eV}$ scenario. As in Fig.~\ref{fig:mid}, the red error bars show a model independent measurement of the neutrino flux in piecewise segments of an $E^{-2}$ spectrum, while the blue error bars and gray upper limits are the corresponding measurements from~\cite{IceCube:2020acn}. The predicted neutrino flux produced by the pure-proton component is conservative, in the sense that only cosmogenic interactions are considered for this population. For a comparison of the expected neutrino flux that additionally allows for interactions of protons in the source environment, we refer the reader to~\cite{Muzio:2023skc}.

\par
The SPL neutrino spectrum, shown in the right panel of Fig.~\ref{fig:hi}, is only marginally hardened by KM3-230213A, making it more difficult to reconcile this model with the data. However, under the UHECR-origin model, a different picture of the neutrino data emerges as shown in the left panel, which is preferred at roughly the $3.3\sigma$ level (see Table~\ref{tab:fitSummaries}). The neutrino data up to roughly \SI{10}{\peta \eV} marks the transition from a low-energy component, which is primarily produced through hadronic interactions of the mixed composition population in their sources, to a UHECR-generated component. In this regime, the data follows to good approximation a SPL, with some deviations from a pure power law visible. Above \SI{100}{\peta \eV}, a significant recovery in the neutrino spectrum is predicted. These neutrinos are produced via interactions of the pure-proton component with the CMB. Under this model, KM3-230213A would be interpreted as the first observed GZK neutrino, suggesting that a wealth of EeV-scale neutrinos will be observed by the next-generation of neutrino detectors.

\begin{figure}[htbp]
\centering
\includegraphics[width=0.9\textwidth]{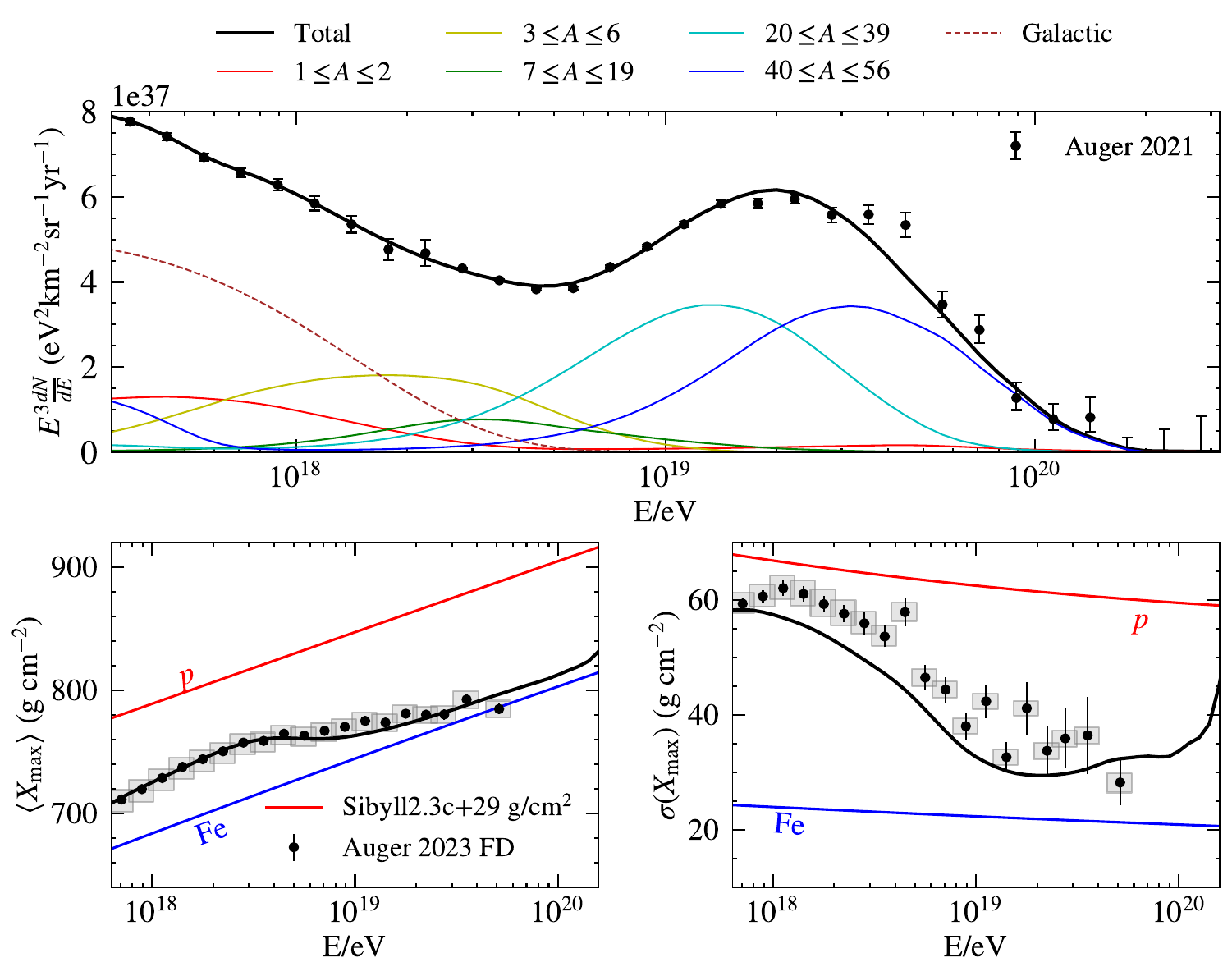}
\caption{Same as Fig.~\ref{fig:fid_CRspec} for the \SI{1}{\exa \eV} KM3NeT/ARCA scenario and with an additional proton component at the highest energies. See Fig.~\ref{fig:hi_CRxmax} for a more detailed comparison of the predicted $X_\mathrm{max}$ distributions to data.} \label{fig:hi_CRspec}
\end{figure}
A summary of the UHECR spectrum and composition data from Auger and our model predictions is shown in Fig.~\ref{fig:hi_CRspec}, with the full $X_\mathrm{max}$ distributions shown in Fig.~\ref{fig:hi_CRxmax} in the \nameref{sec:methods}. The most notable prediction of this model is a recovery of the proton spectrum at Earth above \SI{10}{\exa \eV}. This results in a deviation of the model's predicted $\sigma(X_\mathrm{max})$ from the Auger published data above roughly \SI{20}{\exa \eV}~\cite{PierreAuger:2023kjt}. However, due to the relatively low $X_\mathrm{max}$ statistics in this energy range, the long tail of the Gumbel distribution may not yet be adequately characterized to accurately measure $\sigma(X_\mathrm{max})$. A more detailed picture of the degree to which our model agrees with current data can be obtained from the full $X_\mathrm{max}$ distributions in Fig.~\ref{fig:hi_CRxmax}, which show good agreement between our model and the data. The only significant tension comes from the highest-energy bin, but this is still within uncertainties of the latest proton fractions published by Auger~\cite{PierreAuger:2023xfc}.

\subsection*{\SI{220}{\peta\electronvolt} case: Implications from the best-fit energy}

\begin{figure}[htbp]
    \centering
    \includegraphics[width=0.48\linewidth] {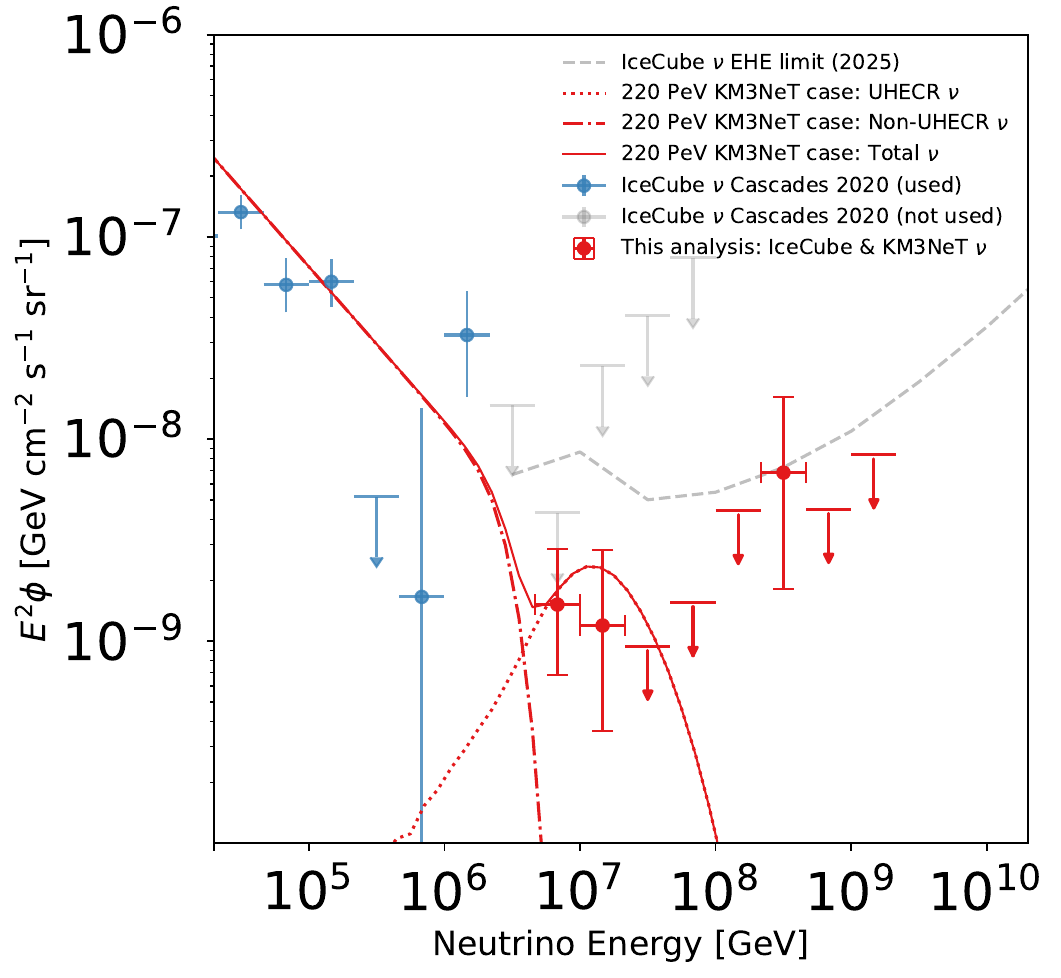} 
    \includegraphics[width=0.48\linewidth]{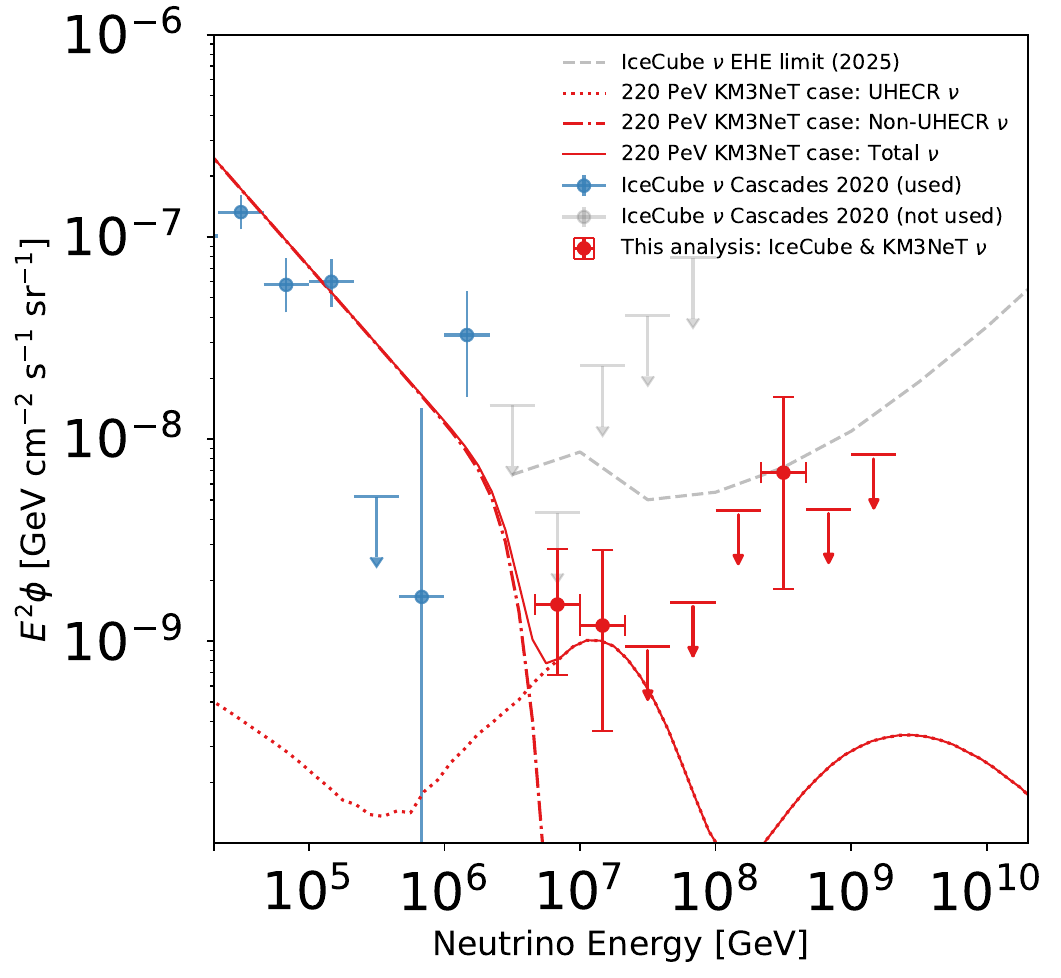} 
    \includegraphics[width=0.48\linewidth] {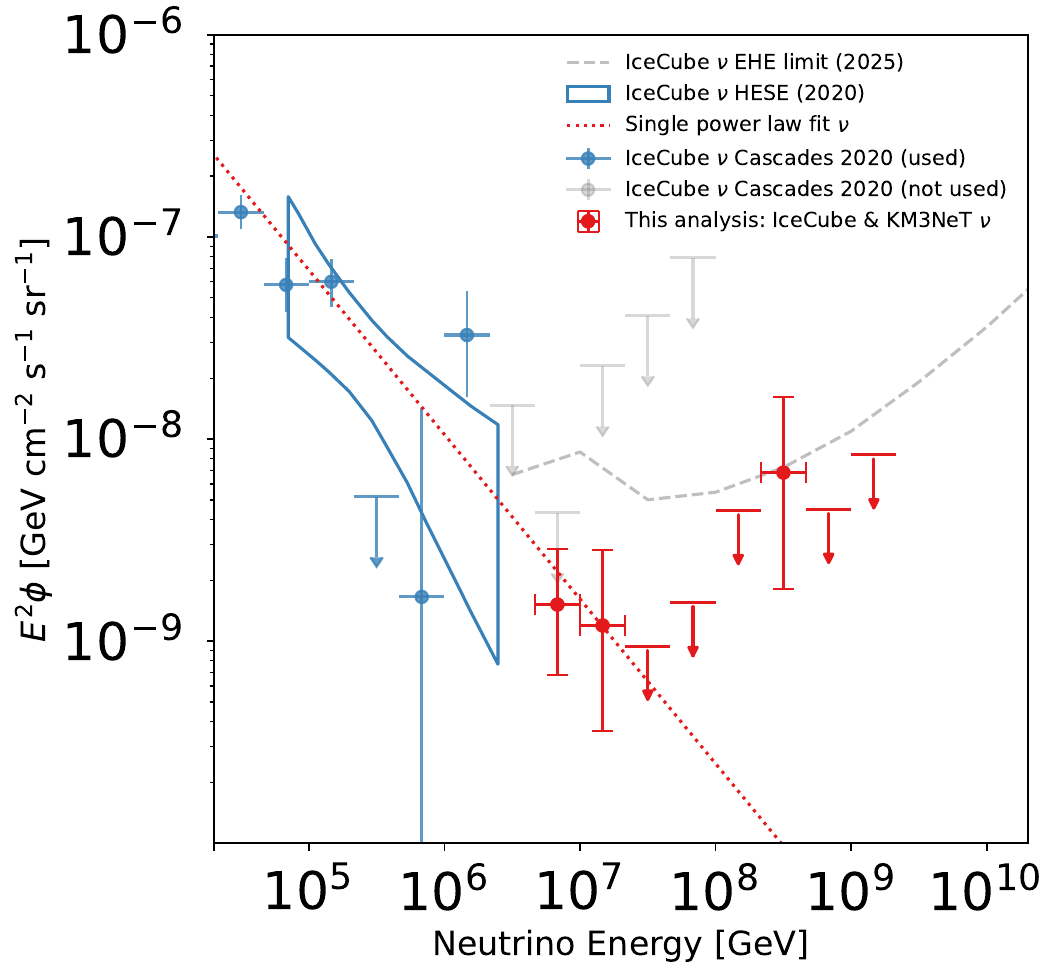} 
    \caption{Same as Fig.~\ref{fig:mid} except KM3-230213A is assumed have an energy of \SI{220}{\peta \eV} and the top panels show UHECR models with (right) and without (left) an extra pure-proton population at the highest energies. The bottom panel shows the best-fit SPL flux model. The bottom panel shows the best-fit SPL flux model. The segmented flux (red points) is re-fit with KM3-230213A at this energy.}
    \label{fig:central}
\end{figure}
\par
In the previous sections, we considered the implications of true neutrino energies which roughly correspond to the $1\sigma$~error on the energy reported by~\cite{KM3NeT:2025npi}. In this section, we consider the case where the true neutrino energy is the central value of \SI{220}{\peta \eV}. 

To fit the neutrino data in this case we consider three models: a SPL flux, a single-population UHECR flux, and a UHECR flux with a contribution due to a second population producing UHE protons. Since the joint UHECR-neutrino deviance is dominated by contributions from the UHECR spectrum and composition (see Table~\ref{tab:fitSummaries}), the assumed \SI{220}{\peta \eV} energy results in only marginal differences compared to the best-fit models presented in the previous sections. A comparison of the results under each of these models is shown in Fig.~\ref{fig:central}.

A \SI{220}{\peta \eV} neutrino would be a significant upward fluctuation of a SPL flux, which is driven by neutrino observations at lower energies. However, this energy is high enough that it would represent an even larger upward fluctuation of the neutrino flux predicted for a single UHECR population. This manifests as a larger total neutrino deviance compared to the SPL model and $p$-value in Table~\ref{tab:fitSummaries} (though this difference is not statistically significant). This tension could be eased by considering functional forms other than a pure blackbody for the photon spectral density in the UHECR source environment. A modified blackbody or broken power-law spectrum would broaden the cutoff of the neutrino spectrum to better describe the data. However, the data can still be well-described by a two-component UHECR model even for the central energy value, being preferred over the SPL model at the $1.5\sigma$~level.

These results indicate that a \SI{220}{\peta \eV} neutrino could be the result of either the high-energy end of a UHECR-produced astrophysical neutrino component or the onset of a cosmogenic component. This ambiguity, which only the next generation of high-energy neutrino observatories will be able to resolve, underscores the sensitivity of the interpretation of the KM3NeT observation to the exact energy of this neutrino. However, in all cases a recovery of the neutrino spectrum above \SI{5}{\peta \eV} describes the highest-energy neutrino at least as well as a SPL model, with the added benefit of also explaining the UHECR spectrum and composition.

\section*{Conclusion and future prospects}\label{sec:Gen2}

In this work, we explored connections between UHECRs and high-energy neutrinos using data from the world’s largest particle astrophysics observatories-the Pierre Auger Observatory, the IceCube Neutrino Observatory, and the KM3NeT/ARCA neutrino telescope. The results, obtained using the latest publicly available datasets, have implications for models of a common origin of UHECRs and neutrinos.

Our analysis reveals the possibility of an emerging neutrino population beyond \SI{5}{\peta \eV}, linked to UHECRs, based on maximum likelihood fits to neutrino and cosmic ray fluxes and cosmic ray $X_\mathrm{max}$ distributions. The UHECR-linked neutrinos exhibit a recovery in energy flux, either around \SI{30}{\peta \eV} or \SI{1}{\exa \eV}, depending on the assumed neutrino energy of the latest KM3NeT/ARCA candidate. The former is dominated by neutrinos directly produced in UHECR sources, while the latter represents cosmogenic neutrinos from a subset of proton-dominated UHECR sources. Combined with the longer-tailed $X_\mathrm{max}$ distributions from Auger, an $E_\nu$ of around \SI{1}{\exa \eV} for KM3-230213A would be the first indications of an as-yet undiscovered source population of UHE protons. It is important to note, however, that the energy of KM3-230213A is not known to high precision and its inference depends on the flux itself~\cite{Yuan:2025isv}. The results presented here should be interpreted as illustrative for a few scenarios that attempt to cover a range of plausible values.

Of course, the best argument that a significant neutrino flux exists above \SI{5}{\peta \eV} will be provided by additional neutrino observations at these energies. Based on current estimates, IceCube and KM3NeT/ARCA should be able to detect a handful more events at the lower end of this energy range in the coming years. However, the next-generation of neutrino observatories (with substantially larger active target volumes~\cite{Chen:2026jdx,TRIDENT:2022hql,Zhang:2024slv,Hymon:2026hcm,TDR}, or new detection techniques~\cite{RNO-G:2020rmc,TAMBO:2025jio,Otte:2019aaf,TDR,GRAND:2018iaj}) will be required in order to distinguish between the SPL expectation and a recovery in the neutrino spectrum.

\begin{figure}[htbp]
\centering
\includegraphics[width=0.75\textwidth]{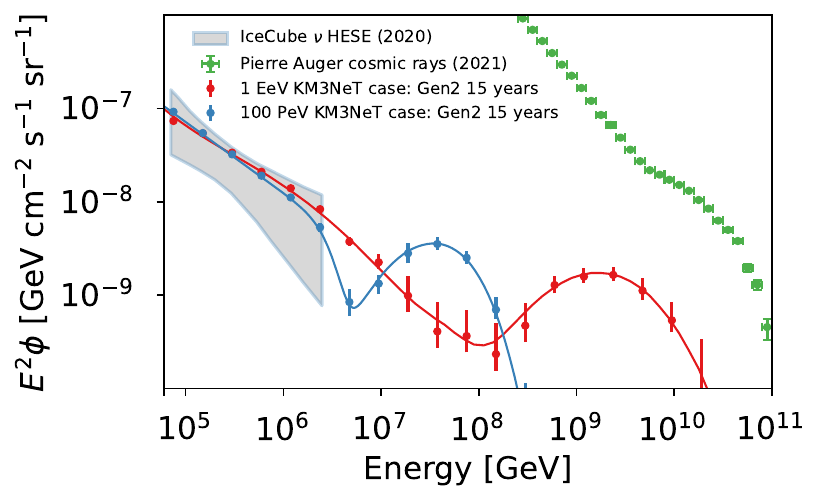}
\caption{Projected neutrino flux from an Asimov test for IceCube-Gen2 under UHECR--$\nu$ models assuming a 100\,PeV (blue) and 1\,EeV (red) neutrino observation. The 1\,EeV model includes an additional pure-proton population at the highest energies. The shaded region indicates the 1$\sigma$ uncertainty from a single power-law fit to IceCube’s HESE data~\cite{IceCube:2020wum}, and the green points represent the cosmic-ray flux from the Pierre Auger Observatory~\cite{PierreAuger:2021hun}.
}\label{fig:gen2}
\end{figure}
Figure~\ref{fig:gen2} shows the potential for one such observatory, the IceCube-Gen2 detector~\cite{TDR}, to discriminate between these possibilities. IceCube-Gen2 has been proposed to expand the instrumented volume of IceCube's optical array from one to at least eight cubic kilometers. It will feature a core in-ice array of optical sensors, each outfitted with multiple PMTs~\cite{IceCube-Gen2:2023ycx}, and a shallow array of radio antennas placed up to $\sim \SI{100}{\m}$ deep in the ice and covering \SI{500}{\kilo \m^2}, significantly enhancing sensitivity to ultrahigh-energy neutrinos at the EeV scale~\cite{IceCube-Gen2:2023vtj}. An Asimov test assuming a common origin model of UHE source neutrinos (blue) and UHE cosmogenic neutrinos (red) is shown in Fig.~\ref{fig:gen2}. The sensitivities shown here assumes 15 years of data collection from both the optical and radio components of IceCube-Gen2. The energy and angular resolutions of the Gen2 detector were taken from the Technical Design Report~\cite{TDR}, and the fitting tool used was the \emph{toise} software package~\cite{vanSanten:2022wss}. If the neutrino flux above \SI{5}{\peta \eV} is, indeed, dominantly produced by UHECRs (in particular through their interactions in sources), IceCube-Gen2 will have the effective volume required to detect and characterize this flux at a statistically significant level.

The AugerPrime upgrade also offers a valuable opportunity to incorporate ground-based measurements of UHECR mass composition~\cite{PierreAuger:2016qzd}, although the primary bottleneck for such joint global-fit analyses remains the limited statistics of neutrino data. After more than a decade of operation, IceCube has detected only three neutrino candidates with energies above 5 PeV. While ARCA eventually will have an effective volume comparable to that of IceCube, significant advancements will require larger target masses and next-generation detectors such as IceCube-Gen2. Future neutrino observatories could provide the necessary data to probe and constrain the properties of UHECR accelerators and their host environments~\cite{Valera:2022wmu}, and perhaps allow for joint UHECR-neutrino searches to pinpoint their sources. 

\section*{Methods}\label{sec:methods}

\par
To fit the Auger spectrum and composition measurements, we adopt the Unger-Farrar-Anchordoqui (UFA) CR source model~\cite{Unger:2015laa} as extended in~\cite{Muzio:2019leu,Muzio:2021zud}. All CR and neutrino flux predictions are calculated using the \textsc{SourceProp} code,\footnote{https://github.com/MichaelUnger/SourceProp} which implements the UFA model. Source emission is modeled through a UHECR spectrum injected into a larger environment host to the accelerator. These CRs interact with the environment's ambient photons and gas (producing secondary nuclei, photons, and neutrinos) while propagating through its turbulent magnetic field. The photodisintegration and spallation interactions in the source generically produce a mixed-composition flux of CRs escaping the source environment. After escaping, UHECRs undergo further photohadronic interactions with the CMB and EBL as they propagate to Earth. Importantly, the UFA model is able to characterize a broad range of source types due to its general source parametrization and lack of absolute scale.  

\par
The injected CR spectrum is modeled as a single-power law spectrum with a cutoff predicted by simulations of particle acceleration through magnetically dominated turbulence~\cite{Comisso:2024ymy}:
\begin{align}\label{eq:CRspectrum}
    J_A(E) \propto E^{\gamma_\mathrm{inj}}~\mathrm{sech}[\left(E/E_{\mathrm{cut},A}\right)^2]~,
\end{align}
where the cutoff energy $E_{\mathrm{cut},A} = Z_A E_{\mathrm{cut},p}$ of each nucleus follows a Peters cycle~\cite{Peters:1961mxb} (composition-independent rigidity cutoff). We approximate the composition of the accelerated spectrum through five nuclei (representing $p$, He-like, CNO-like, Si-like, and Fe-like mass groups) with relative abundances as free parameters of the model. In total the injected CR spectrum is characterized through $7$~free parameters. 

\par
The source environment is characterized through $5$~free parameters which (indirectly) control the photon-to-gas density, relative magnetic field strength and coherence length, and relative source size. These parameters are discussed in-depth in~\cite{Muzio:2021zud,Muzio:2022bak}. Sources are assumed to be standard, modulo their relative luminosity-density which is captured through a model of their cosmological evolution. For simplicity, we assume this evolution follows the star formation rate (SFR) evolution~\cite{Robertson:2015uda} up to a maximum comoving distance of \SI{4.2}{\giga pc}.

\par
A single-power law Galactic CR spectrum (with exponential cutoff) is added to the predicted CR spectrum as a nuisance component to model the Galactic-to-extragalactic transition. For simplicity, we approximate this component as having a pure composition. The Galactic nuisance component adds $4$~free parameters.

\par
Extragalactic propagation is modeled using pre-tabulated propagation matrices generated using \textsc{CRPropa3}~\cite{AlvesBatista:2016vpy} assuming EBL model from~\cite{Gilmore:2011ks} and photodisintegration cross-sections from \textsc{Talys v1.8}~\cite{Koning:2005ezu}.

\par
In the case of the $1$~EeV KM3-230213A scenario, we consider one additional extragalactic CR component: a pure-proton component with an escaping spectrum parametrized by a single-power law and exponential cutoff. This component is assumed to come from a second source population (given their much larger rigidity cutoff) with few in-source interactions. For simplicity we assume this population also follows a SFR evolution. This second proton component adds $3$~additional free parameters to the model.

\par
The neutrino flux is modeled through two components. The first component, which dominates at high energies, is the neutrino flux produced by UHECRs inside sources and through extragalactic propagation. This component is fully captured by the model described above and adds no additional free parameters. Its normalization is set by the fit of the propagated CR flux to the Auger spectrum. Because the neutrino flux receives contributions from sources out to high redshift while the observed UHECRs originate within a few hundred Mpc, the ratio of the neutrino to the CR flux depends on the assumed source evolution. Flavors are tracked separately for neutrinos from $\pi^\pm$ decay and $\bar{\nu}_e$ from neutron $\beta$-decay, and propagated to Earth with flavor-averaged oscillation probabilities~\cite{Anchordoqui:2013dnh}.

The second component, which serves as a nuisance component to capture the low-to-high energy transition, is modeled as a single-power law with a cutoff following the same functional form as that in~\eqref{eq:CRspectrum} (with an assumed $1:1:1$~flavor ratio). This low-energy neutrino nuisance component adds $3$~free parameters.

\begin{table}[]
    \centering
    \begin{tabular}{lllll}
        \toprule
        \textbf{Parameter} & \textbf{\SI[text-series-to-math]{100}{\textbf{\peta\eV}}} & \textbf{\SI[text-series-to-math]{220}{\textbf{\peta\eV}}} & \textbf{\SI[text-series-to-math]{220}{\textbf{\peta\eV}}$^*$} & \textbf{\SI[text-series-to-math]{1}{\textbf{\exa\eV}}$^*$} \\
        \hline\midrule
        $\gamma_{\mathrm{inj}}$ & -2.27 & -2.26 & -2.27 & -2.27 \\
        $\log_{10}(R_\mathrm{max}/\mathrm{V})$ & 18.67 & 18.67 & 18.67 & 18.68 \\
        $\log_{10}r_{\mathrm{esc}}$ & 2.00 & 1.99 & 1.97 & 1.97 \\
        $f_g$ & 0.00 & 0.00 & 0.00 & 0.00 \\
        $\log_{10}(R_\mathrm{diff}/\mathrm{V})$ & 15.80 & 16.08 & 16.02 & 16.01 \\
        $\tanh(\log_{10} r_\mathrm{size})$ & 1.00 & 1.00 & 1.00 & 1.00 \\
        $T/\mathrm{K}$ & 715.35 & 721.95 & 721.40 & 721.60 \\
        $(A_p$, $f_p)$ & (1, 0.0\%) & (1, 0.0\%) & (1, 0.0\%) & (1, 0.0\%) \\
        $(A_\mathrm{He}$, $f_\mathrm{He})$ & (3, 11.3\%) & (3, 11.8\%) & (3, 11.9\%) & (3, 11.9\%) \\
        $(A_\mathrm{CNO}$, $f_\mathrm{CNO})$ & (7, 39.5\%) & (7, 39.6\%) & (7, 39.4\%) & (7, 39.3\%) \\
        $(A_\mathrm{Si}$, $f_\mathrm{Si})$ & (24, 4.7\%) & (24, 4.2\%) & (24, 4.3\%) & (24, 4.3\%) \\
        $(A_\mathrm{Fe}$, $f_\mathrm{Fe})$ & (52, 44.4\%) & (52, 44.5\%) & (52, 44.5\%) & (52, 44.5\%) \\
        $\gamma_{\mathrm{gal}}$ & -2.86 & -2.87 & -2.87 & -2.87 \\
        $\log_{10}(E_\mathrm{max}^\mathrm{gal,Fe}/\mathrm{eV})$ & 18.17 & 18.17 & 18.17 & 18.17 \\
        $f_\mathrm{gal}$ & 0.60 & 0.60 & 0.60 & 0.60 \\
        $A_\mathrm{gal}$ & 44 & 44 & 44 & 44 \\
        $\gamma_\nu$ & -2.77 & -2.77 & -2.77 & -2.77 \\
        $\log_{10}(E_{\mathrm{max},\nu}/\mathrm{eV})$ & 15.37 & 15.36 & 15.37 & 15.37 \\
        $\phi_{\nu 0}$ [$10^{-18}$/(GeV~cm$^2$~s~sr)] & 5.14 & 5.14 & 5.14 & 5.14 \\
        $\gamma_{\mathrm{UHE}p}$ & N/A & N/A & -2.20 & -2.18 \\
        $\log_{10}(E_\mathrm{max}^{\mathrm{UHE}p}/\mathrm{eV})$ & N/A & N/A & 21.00 & 21.00 \\
        $f_{\mathrm{UHE}p}$ & N/A & N/A & 0.05 & 0.05 \\
        \midrule
        $\gamma_\mathrm{SPL}$ & -2.82 & -2.81 & -2.81 & -2.81 \\
        $\phi_\mathrm{SPL}$ [$10^{-18}$/(GeV~cm$^2$~s~sr)] & 4.97 & 4.97 & 4.97 & 4.97 \\
        \bottomrule
    \end{tabular}
    \caption{Model parameters for the UHECR models presented in the main text. Models are labelled according to the assumed energy for the neutrino observed by KM3NeT. Models with an asterisk include an additional proton component. The last two rows give the parameters of the SPL model of each scenario parametrized by Eq.~\eqref{eq:SPL}.
    }
    \label{tab:modelParams}
\end{table}

\par
In total, the baseline UHECR source model has $19$~free parameters. If a second proton component is considered, there are $22$~total free parameters. The values of these parameters for the models presented in the main text are given in Table~\ref{tab:modelParams}. A detailed discussion of the connection between these model parameters and astrophysical source properties can be found in~\cite{Muzio:2022bak}. Several parameters are defined with respect to a reference nucleus, which is taken to be \SI{10}{\exa \eV} Fe$^{56}$. Briefly, these parameters are defined as follows: $\gamma_\mathrm{inj}$ is the power-law index, $E^{\gamma_\mathrm{inj}}$, of the accelerated CR spectrum; $R_\mathrm{cut} = E_{\mathrm{cut},A}/Z$ is the rigidity scale of the cutoff in~\eqref{eq:CRspectrum}; $r_\mathrm{esc}$ is the ratio of the escape-to-interaction times for the reference nucleus; $f_g$ is the fraction of interactions which are hadronic for the reference nucleus; $R_\mathrm{diff}$ is the rigidity scale of the magnetic field, assumed to be turbulent with a Kolmogorov spectrum; $r_\mathrm{size}$ is the ratio of the effective source size to the magnetic field's coherence length; $T$ is the blackbody temperature of the ambient photon field in the source; $(A_i,f_i)$ are the mass numbers and composition fractions injected into the source (representative of mass group $i$), where $f_i$ is defined at a fixed energy $E \ll E_{\mathrm{cut},A}$ for all $A$; $\gamma_\mathrm{gal}$ is the power law index, $E^{\gamma_\mathrm{gal}}$ of the Galactic component; $E_\mathrm{max}^\mathrm{gal,Fe}$ is the maximum energy of Galactic iron, where the Galactic component is cutoff exponentially, which sets the maximum rigidity of the Galactic component; $A_\mathrm{gal}$ is the mass assumed for the Galactic component; $\gamma_\nu$ is the power law index, $E^{\gamma_\nu}$, of the low-energy neutrino component; $E_{\mathrm{max},\nu}$ is the low-energy neutrino component's cutoff energy scale; $\phi_{\nu0}$ is the normalization of the low-energy neutrino spectrum at \SI[parse-numbers = false]{10^{14}}{\eV}; $\gamma_{\mathrm{UHE}p}$ is the power law index, $E^{\gamma_{\mathrm{UHE}p}}$ of the second proton component; $E_\mathrm{max}^{\mathrm{UHE}p}$ is the energy of the second proton component's exponential cutoff; and, $f_{\mathrm{UHE}p}$ is the fraction of the total energy above \SI[parse-numbers = false]{10^{17}}{\eV} escaping both source populations carried by the second proton component. 

\par
The last two rows of Table~\ref{tab:modelParams} list the spectral index $\gamma_\mathrm{SPL}$ and the all-flavor normalization $\phi_\mathrm{SPL}$ at \SI[parse-numbers = false]{10^{14}}{\eV} of the SPL model, Eq.~\eqref{eq:SPL}, which is fit to the neutrino data alone and is independent of the UHECR model. Because the normalization and index of the SPL are constrained predominantly by the IceCube flux measurements below \SI{2}{\peta \eV}, its parameters are nearly independent of the assumed energy of KM3-230213A: $\phi_\mathrm{SPL}$ is unchanged and $\gamma_\mathrm{SPL}$ hardens by only ${\simeq}0.01$ between the \SI{100}{\peta \eV} and \SI{1}{\exa \eV} scenarios. The dependence of $D_{\nu,\mathrm{hi}}$ and of the $p$-values in Table~\ref{tab:fitSummaries} on the assumed energy is therefore set by how well this nearly fixed SPL accommodates the KM3NeT event.

\par
These parameters are fit to maximize the joint likelihood to the UHECR spectrum \& composition and neutrino data. In practice, this is carried out by minimizing the deviance $D=-2\ln(L/L_\mathrm{sat})$, where $L$ is the model likelihood and $L_\mathrm{sat}$ is the likelihood of a model that perfectly describes the data. The total deviance used in the minimization of the UHECR model is given by $D=D_{\mathrm{CR},J} + D_{\mathrm{CR},X} + D_{\nu,\mathrm{lo}} + D_{\nu,\mathrm{hi}}$\footnote{The SPL and segmented neutrino flux models are constrained by the $D_{\nu,\mathrm{lo}} + D_{\nu,\mathrm{hi}}$ term alone.}.

\par
The UHECR spectrum is fit using a saturated Poisson likelihood 
\begin{align}
    D_{\mathrm{CR},J} = 2\sum_i m_i - d_i + d_i\ln(d_i/m_i)~,
\end{align}
where the sum is over energy bins $17.5 \leq \lg(E_i/\mathrm{eV}) \leq 22$, $d_i$ is the observed number of events in bin $i$, and $m_i$ is the predicted number of events by the model. Spectrum data points are taken from~\cite{PierreAuger:2021hun}.

\begin{figure}[htbp]
\centering
\includegraphics[width=0.9\textwidth]{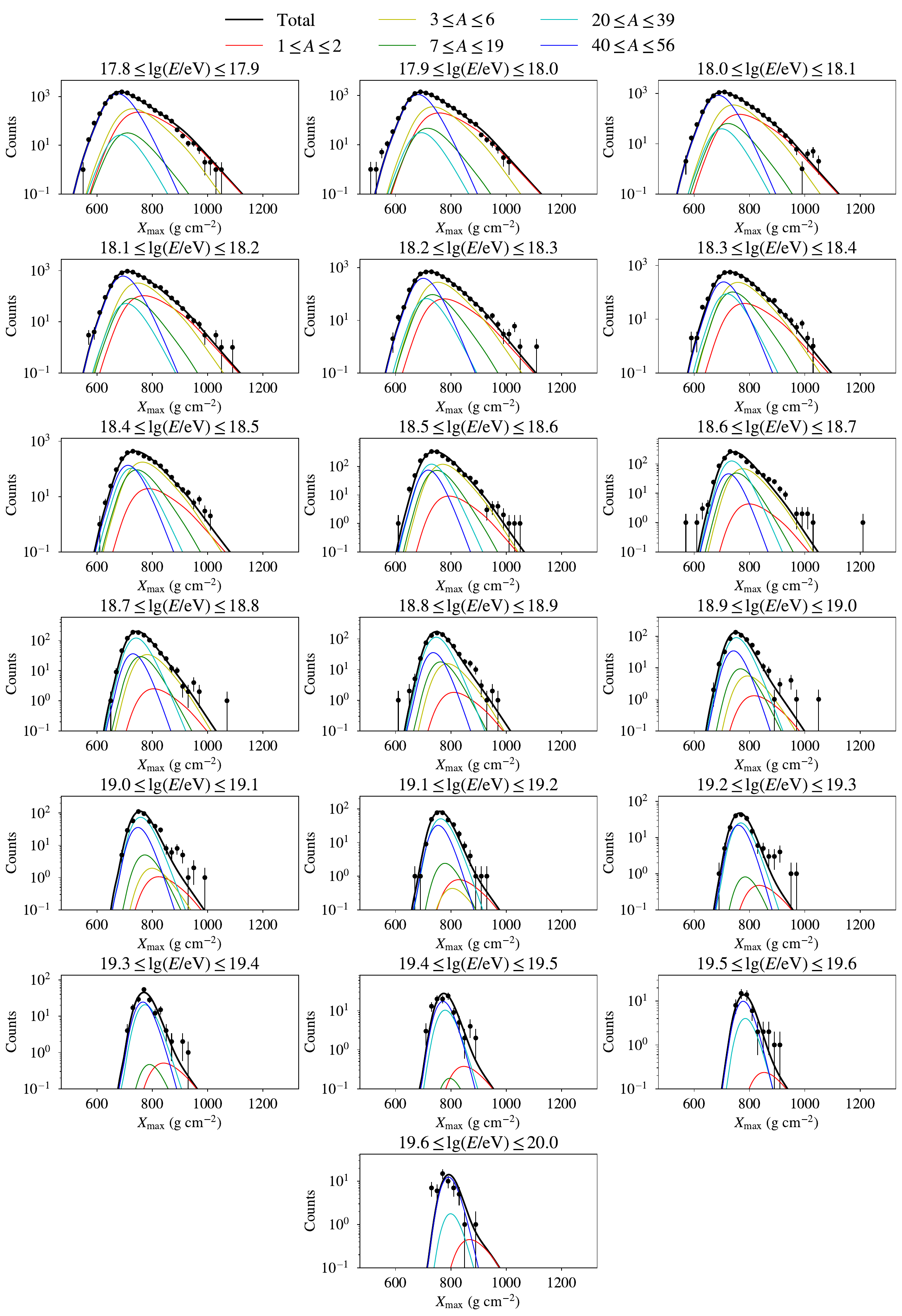}
\caption{Predicted $X_\mathrm{max}$ distribution for the best-fit model with an additional high energy proton component, assuming KM3-230213A has an energy of \SI{1}{\exa\eV}. The total forward-folded distribution (black lines) is compared to Auger measurements~\cite{PierreAuger:2023kjt}, and broken down by mass group contribution (colored lines).}\label{fig:hi_CRxmax}
\end{figure}

\par
Following~\cite{PierreAuger:2022atd}, the UHECR composition data (namely, the depth of shower maximum distributions) are fit using a multinomial likelihood
\begin{align}
    D_{\mathrm{CR},X} = 2\sum_{i,j} d_{ij} \ln\left(\frac{d_{ij}}{n_i p_{ij}}\right)~,
\end{align}
where the sum is over energy bins $17.8 \leq \lg(E_i/\mathrm{eV}) \leq 20$ and depth of shower maximum bins $500~\mathrm{g/cm}^{2} \leq X_\mathrm{max,j} \leq 1300~\mathrm{g/cm}^2$, $n_i$ is the total number of observed events in the $i$th bin's $X_\mathrm{max}$ distribution, $d_{ij}$ is the number of observed events in the $j$th $X_\mathrm{max}$ bin, and $p_{ij}$ is the model probability for a shower in energy bin $i$ to have $X_\mathrm{max}$ falling in the $j$th bin. $p_{ij}$ is given by a sum of probability distributions $G^A_{\mathrm{reco},ij}$ weighted by the fraction of the total model flux contributed by CRs of mass $A$. These probability distributions are calculated by forward folding the generalized Gumbel distribution $G^A_{ij}$ for a CR with mass $A$ and energy $E_i$ with the Auger acceptance and resolution~\cite{PierreAuger:2014sui}. We use the parameters from~\cite{Arbeletche:2019qma} for \textsc{Sibyll2.3c}~\cite{Fedynitch:2018cbl} for the true generalized Gumbel distributions, adding an additional $+29$~g/cm$^2$ to the distribution based on~\cite{PierreAuger:2024neu}. The $X_\mathrm{max}$ data points are taken from~\cite{PierreAuger:2023kjt}. An example of these distributions, along with their model fits, is shown for the \SI{1}{\exa\eV} UHECR model case in Fig.~\ref{fig:hi_CRxmax}.

\par
The neutrino likelihood is split into a low- and high-energy contribution. At lower energies, where more statistics are available, it becomes important but challenging to model the background, detector and systematic uncertainties. Thus, we fit to the published neutrino flux directly as that includes these effects to the best of our knowledge at the time. We use the best-fit piecewise differential flux as measured by IceCube in Ref.~\cite{IceCube:2020acn}, converted to an all-flavor flux assuming a $1$:$1$:$1$ flavor ratio. The measured flux in each piecewise segment has asymmetric errors (or upper bounds) that contain both statistical and systematic uncertainties. In order to incorporate these measurements into a likelihood, we assume the measured flux in each segment can be modeled by a gamma distribution whose mode is set to the published best-fit point. This is motivated by interpreting the gamma distribution as a continuous analog of the negative binomial distribution, which itself generalizes the Poisson distribution to allow for a variance that differs from the mean and hence can model additional uncertainties on top of pure Poisson statistics. The gamma log-likelihood is then given by 
\begin{align}
    D_{\nu,\mathrm{lo}} = 2 \sum_i (\alpha_i-1) \ln(y_i/\phi_i) + (\phi_i - y_i)/\theta_i~,
\end{align}
where the sum is over energy bins $\SI{30}{\tera \eV} \leq E_i \leq \SI{2}{\peta \eV}$, $y_i$ is the central value of the piecewise flux in that bin, and $\phi_i$ is the predicted all-flavor flux under our model. The parameters $\alpha_i$ and $\theta_i$ correspond to the shape and scale parameters of the underlying gamma distribution, and are fixed by the best-fit point and its associated uncertainty. That is, for a given energy bin $i$, $\alpha_i$ and $\theta_i$ are obtained by solving the following system of equations,
\begin{equation}
    \begin{cases}
    y_i = (\alpha_i-1)\theta_i \\
    \int_{y_i-\sigma^-_i}^{y_i+\sigma^+_i} \mathrm{Gamma}(x; \alpha_i, \theta_i) dx = 0.68~,
    \end{cases}
\end{equation}
where $\sigma_i^\pm$ are the upper and lower error bars reported on $y_i$ and $\mathrm{Gamma}(x; \alpha_i, \theta_i)$ is the probability density function of the gamma distribution. In the case of a reported upper limit, we take $y_i = \sigma_i^-=0$. We note that correlations across segments may exist, but are not publicly available. As the sample is dominated by charged-current electron and tau neutrino interactions in which the reconstructed energy is a good proxy for the neutrino energy, we assume the correlations are negligible.

At neutrino energies above \SI{5}{\peta \eV}, where only four neutrino events have been observed by both IceCube and KM3NeT, we again use a saturated Poisson likelihood. The deviance is given by
\begin{align}\label{eq:Lnuhi}
    D_{\nu,\mathrm{hi}} = 2 \sum_{E_i \leq 1~\mathrm{EeV}} \sum_{a,k,\rho} m_{aik\rho} - d_{aik\rho} + d_{aik\rho}\ln(d_{aik\rho}/m_{aik\rho})
    + 2 \sum_{E_i > 1~\mathrm{EeV}} m_i~.
\end{align}
Below \SI{1}{\exa \eV}, $a$ indexes the neutrino datasets given in Table~\ref{tab:nuDatasets}, $\rho$ the neutrino flavor, $i$ the neutrino energy, $k$ the $\cos\theta_k$ zenith bins, and $d$ and $m$ represent the observed and predicted number of events in a bin, respectively. To calculate $m_{aik\rho}$, the predicted single-flavor flux $\phi_{\nu_\rho + \overline{\nu}_\rho}$ is multiplied by the exposure of dataset $a$ to flavor $\rho$ neutrinos and integrated over the area of the $ik$th energy-$\cos\theta$ bin. Note that this assumes an isotropic flux. As described in the main text and listed, three IceCube, one KM3NeT/ARCA, and two Auger datasets are included. The effective areas for these datasets were extrapolated to $1$~EeV where necessary.

The requirement $E_i\leq \SI{1}{\exa \eV}$ is applied to the bin centers, and KM3-230213A is ensured to fall into that portion of the deviance calculation for all three energy scenarios. Above \SI{1}{\exa \eV}, the non-observation of neutrinos is modeled by calculating the total number of neutrinos $m_i$ predicted in the $i$th energy bin when multiplied to the latest IceCube exposure from Ref.~\cite{IceCubeCollaborationSS:2025jbi,DVN/JHK49D_2025}.

\par
For visualization purposes, we also calculate a segmented flux, shown as red points in Figs.~\ref{fig:mid},~\ref{fig:hi}, and~\ref{fig:central}, obtained by minimizing the first term in Eq.~\eqref{eq:Lnuhi} without the requirement that $E_i \leq \SI{1}{\exa \eV}$. The parametric flux is replaced by a piecewise $E^{-2}$ model~\cite{IceCube:2020wum,IceCube:2020acn} with eight energy segments of width $\Delta\log_{10}E_\nu = 1/3$, over the range $10^{6.67}$--$10^{9.33}$~GeV (\SI{4.6}{\peta \eV} to \SI{2.2}{\exa \eV}). The flux normalization in each segment is simultaneously fit over the observable space derived from each dataset's effective area. No cosmic-ray or low-energy neutrino data are included in this fit. The expected counts are computed from the published effective areas of each dataset over their tabulated energy and zenith ranges. A $1:1:1$ flavor ratio is assumed. Error bars show the uncertainty on the flux in each segment, obtained from evaluating the \SI{68}{\%} profile-likelihood confidence interval for that segment, with other normalizations treated as nuisance parameters. For segments whose best-fit normalization is zero, the (one-sided) $\SI{84}{\%}$ upper bound is shown. The segmented fit is repeated for each  KM3-230213A energy scenario.

Finally, to quantify the UHECR model's ability to describe the neutrino data above \SI{5}{\peta \eV} relative to a single power-law, we calculate the empirical $p$-value of the test statistic,
\begin{align}
    \mathrm{TS} = D_{\nu, \mathrm{UHECR}} - D_{\nu, \mathrm{SPL}}~,
\end{align}
where $D_{\nu} = D_{\nu, \mathrm{hi}}$ is the high-energy neutrino deviance. With this definition, a smaller TS indicates that the UHECR model describes the neutrino data better than the single power-law. 

To determine the TS distribution, we perform pseudo-experiments drawing neutrino events above \SI{5}{\peta \eV} from a Poisson distribution, while leaving the low-energy and CR data fixed. The expected number of neutrinos is determined by folding the best-fit single power-law spectrum with the exposure of the IceCube Northern Tracks, HESE, and PEPE datasets, as well as the KM3NeT/ARCA and the Auger DG and ES datasets. For each realization of the high-energy neutrino data, both the UHECR model and single-power law models are fit, their deviances are calculated, and a TS for the realization is recorded. During this fit, each model is seeded with its best-fit parameters to the real dataset, and parameters are re-optimized over the same ranges. Since the UHECR data dominates the total deviance and is unchanged, the parameters of the UHECR model are essentially unchanged in this process. The single-power law model and low-energy nuisance component of the UHECR component re-optimize their parameters over the range $\gamma_\nu, \gamma_\mathrm{SPL} \in [-4, -2]$, $\phi_{\nu 0}, \phi_\mathrm{SPL} \in [0.317, 317] \times 10^{-18}/($GeV~cm$^2$~s~sr$)$, and, in the case of the nuisance component, $\log_{10}(E_{\mathrm{max},\nu}/\mathrm{eV}) \in [14,19]$. With this TS distribution, we calculate the empirical $p$-value as
\begin{align}
    p = \Pr(\mathrm{TS} \leq \mathrm{TS}_0~| ~H_0)~,
\end{align}
where $\mathrm{TS}_0$ is the TS calculated for the real neutrino data and $H_0$ is the single power-law null hypothesis.

\backmatter



\bmhead{Acknowledgements}

We acknowledge the use of code created by Michael Unger under the EU-funded Grant PIOF-GA2013-624803 and further developed by M.M.~. We thank Albrecht Karle and Michael Unger for reading through an earlier version of the manuscript and providing valuable comments. 

\section*{Author Contribution}
M.M., T.Y., and L.L. wrote and reviewed the manuscript.

\section*{Funding}
M.M. acknowledges support from NSF MPS-Ascend Postdoctoral Award \#2138121 and the John Bahcall Fellowship at the Wisconsin IceCube Particle Astrophysics Center at the University of Wisconsin--Madison. T.Y. and L.L are supported in part by NSF grant PHY-2209445 and by the University of Wisconsin Research Committee with funds granted by the Wisconsin Alumni Research Foundation.

\section*{Data Availability}
The datasets generated during this study may be available from the corresponding author upon reasonable request.

\bibliography{main}{} 
\bibliographystyle{bst/sn-nature}

\end{document}